\documentclass{aa}  

\usepackage{txfonts}
\usepackage{graphicx}
\usepackage[font=small,labelfont=bf]{caption} 
\usepackage{subcaption}
\usepackage{float}
\usepackage{natbib}
\usepackage{amsmath}
\usepackage{url}
\usepackage[breaklinks=true,pdfmenubar=true,pdftoolbar=true,colorlinks=true,citecolor=blue,linkcolor=blue,urlcolor=blue]{hyperref} 
\bibpunct{(}{)}{;}{a}{}{,}

\begin{document}

   \title{APEX and NOEMA observations of H$_{2}$S in nearby luminous galaxies and the ULIRG Mrk~231}
   \subtitle{Is there a relation between dense gas properties and molecular outflows?}

   \author{M.T. Sato
          \inst{1}
          \and
          S. Aalto\inst{1}
          \and
          K. Kohno
          \inst{2}
          \and
          S. K{\" o}nig
          \inst{1}
          \and
          N. Harada
          \inst{3}
          \and
          S. Viti
          \inst{4,5}
          \and
          T. Izumi
          \inst{3}
          \and
          Y. Nishimura
          \inst{2,3}
          \and
          M. Gorski
          \inst{1}
          }

   \institute{Department of Space, Earth and Environment, Chalmers University of Technology, 
   Onsala Space Observatory, SE-439 92 Onsala, Sweden\\
     		\email{mamiko@chalmers.se}
     		\and
             Institute of Astronomy, Graduate School of Science, The University of Tokyo, 
             2-21-1 Osawa, Mitaka, Tokyo 181-0015, Japan\\
          	\and
          	National Astronomical Observatory of Japan, 2-21-1 Osawa, Mitaka, Tokyo 181-8588, Japan\\
			\and
    		Leiden Observatory, Leiden University, PO Box 9513, 2300 RA, Leiden, The Netherlands\\
    		\and
    		Department of Physics and Astronomy, UCL, Gower Street, London, WC1E 6BT, UK\\
             }

   \date{Received XXX; accepted XXX}

 
  \abstract
   {In order to understand the evolution and feedback of Active Galactic Nuclei (AGN) and star formation it is important to use molecular lines as probes of physical conditions and chemistry.}
   {We use H$_{2}$S to investigate the impact of starburst and AGN activity on the chemistry of the molecular interstellar medium in luminous infrared galaxies. Specifically our aim was to search for evidence of shock-enhancement of H$_{2}$S related to galactic-scale mechanical feedback processes such as outflows.}
   {Using the APEX single dish telescope, we have observed the $1_{10}$--$1_{01}$ transition of ortho-H$_{2}$S at 168 GHz towards the centres of twelve nearby luminous infrared galaxies. We have also observed the same line towards the ultra luminous infrared galaxy (ULIRG) Mrk~231 with the NOEMA interferometer.}
   {We have detected H$_{2}$S towards NGC~253, NGC~1068, NGC~3256, NGC~4418, NGC~4826, NGC~4945, Circinus, M~83 and Mrk~231. Upper limits were obtained for NGC~1097, NGC~1377 and IC~860. We also detected line emission from HCN 2--1 in all galaxies in the APEX survey, and HCO$^+$, HNC, CH$_3$CN, CH$_3$OH, H$_2$CS, HOC$^+$ and SO in several of the sample galaxies. Mrk~231 has a rich 2~mm molecular spectrum and, apart from H$_{2}$S, we detect emission from HC$_3$N, CH$_3$OH, HC$^{18}$O$^{+}$, C$_2$S and CH$_3$CCH. Four galaxies show elevated H$_{2}$S emission relative to HCN: Circinus, NGC~3256, NGC~4826 and NGC~4418.  We suggest that the high line ratios are caused by elevated H$_{2}$S abundances in the dense gas. However, we do not find any clear connection between the H$_{2}$S/HCN line intensity ratio, and the presence (or speed) of molecular outflows in the sample galaxies. Therefore H$_{2}$S abundances do not seem to be globally affected by the large-scale outflows. In addition, the H$_{2}$S/HCN line ratio  is not enhanced in the line wings compared to the line core in Mrk~231. This suggests that H$_{2}$S abundances are not increased in the dense gas in the outflow. However, we do find that the H$_{2}$S and HCN luminosities ($L_{\mathrm{H_{2}S}}$ and $L_{\mathrm{HCN}}$) correlate well with the total molecular gas mass in the outflow, $M_{\rm outflow}$(H$_2$), in contrast to $L_{\mathrm{CO}}$ and $L_{\mathrm{HCO^{+}}}$. We also find that the line luminosity of H$_{2}$S correlates with total infrared luminosity in a similar way that H$_{2}$O does.}
   {We do not find any evidence of H$_{2}$S abundance enhancements in the dense gas due to galactic-scale outflows in our sample galaxies, or in the high-resolution study of Mrk~231. We discuss possible mechanisms behind the suggested H$_{2}$S abundance enhancements in NGC~4418, Circinus, NGC~3256 and NGC~4826. These include radiative processes (for example X-rays or cosmic-rays) or smaller scale shocks. Further high resolution, and multi-transition studies are required to determine the cause behind the elevated H$_{2}$S emission in these galaxies.  We suggest that $L_{\mathrm{H_{2}S}}$ serves as a tracer of the dense gas content, similar to $L_{\mathrm{HCN}}$, and that the correlation between $L_{\mathrm{H_{2}S}}$ and $M_{\rm outflow}$(H$_2$) implies a relation between the dense gas reservoir and the properties and evolution of the molecular feedback. This potential link requires further study since it holds important keys to our understanding of how the properties of molecular outflows relate to that of their host galaxies. Finally, the similar IR-correlation coefficients between H$_{2}$S and H$_2$O may indicate that they originate in the same regions in the galaxy: warm gas in shocks or irradiated by star formation or an AGN.}


   \maketitle
%

\section{Introduction}
\label{s:introduction}

\begin{table*}[!h]
	\caption{Basic information of sources}             
	\label{tab:Basic}      
	\centering                          
	\begin{tabular}{l c c c c c c c c c}        
		\hline\hline                 
		Galaxy	&${\alpha}$ (J 2000)& ${\delta}$ (J 2000)	&$D$\tablefootmark{a}& $L_{\mathrm{IR}}$\tablefootmark{b} & type & Outflow\tablefootmark{c} 				& Water vapor\tablefootmark{d} &$t_{\mathrm{int}}$\tablefootmark{e} & rms[LSB/USB]\tablefootmark{f} \\
				& (h:m:s) 		& ($^{\circ}$ : $\,^{\prime}$:$\,^{\prime\prime}$) & (Mpc) & ($L_{\odot}$) & & & & (hours) & (mK)\\   
		\hline 
		\\
		NGC~253	& 00:47:33.1 	& $-$25:17:18	& \phantom{0}3.1		& $3.0\times10^{10}$ 			& SB 		&  \phantom{00}Yes\tablefootmark{(1)}	& No					&1.5	   & 0.48/0.87\\
		NGC~1068& 02:42:40.7 	& $-$00:00:48	& 13.7		& $2.0\times10^{11}$	 		& AGN+SB 	&  \phantom{00}Yes\tablefootmark{(2)}	& No&3.5 & 0.35/0.60\\
		NGC~1097& 02:46:19.0 	& $-$30:16:30	& 16.8					& $5.0\times10^{10}$ 			& LLAGN 	& No									& No&1.2 & 0.30/0.51\\
		NGC~1377& 03:36:39.1 	& $-$20:54:08	& 23.66					& $1.3\times10^{10}$ 			& AGN? 		&  \phantom{00}Yes\tablefootmark{(3)}	& No&7.7 & 0.24/0.60\\
		NGC~3256& 10:27:51.3 	& $-$43:54:13	& 35.35				 	& $3.6\times10^{11}$ 			& SB 		&  \phantom{00}Yes\tablefootmark{(4)}	&  \phantom{0}Yes\tablefootmark{(1)} &6.9 & 0.20/0.38\\
		NGC~4418& 12:26:54.6 	& $-$00:52:39	& 31.9	 	& $1.0\times10^{11}$ 			& AGN 		&  \phantom{00}Yes\tablefootmark{(5)}	&  \phantom{0}Yes\tablefootmark{(2)} &1.1 & 0.41/0.67\\
		Mrk~231	& 12:56:14.2 	& $+$56:52:25	& 171.8		& $3.2\times10^{12}$ \phantom 	& AGN+SB	&  \phantom{00}Yes\tablefootmark{(6)} 	&  \phantom{0}Yes\tablefootmark{(3)} &&0.6(mJy/beam)/ -\\
		NGC~4826& 12:56:43.6 	& $+$21:40:59	& \phantom{0}3.1	 	& $4.0\times10^{9}$ \phantom 	& AGN		& No									& No &2.8 & 0.27/0.66\\
		NGC	~4945& 13:05:27.5 	& $-$49:28:06 	& \phantom{0}3.9		& $3.0\times10^{10}$ 			& AGN/SB	&  \phantom{00}Yes\tablefootmark{(7)}	& No &4.9 & 0.46/1.0\\
		IC~860	& 13:15:03.5	& $+$24:37:08	& 59.1		& $1.5\times10^{11}$ 			& LIRG 		&  \phantom{00}Yes\tablefootmark{(8)}	&  \phantom{0}Yes\tablefootmark{(4)}&2.6 & 0.52/0.80\\
		NGC	~5128& 13:25:27.6 	& $-$43:01:09	& \phantom{0}4.0	 	& $2.0\times10^{11}$ 			& AGN 		&  \phantom{00}Yes\tablefootmark{(9)}	& No&1.4 & -~~~~/~~~~-\\
		M~83	& 13:37:00.9 	& $-$29:51:56	& \phantom{0}3.6	 	& $1.0\times10^{10}$ 			& SB 		& No									& No&1.4 & 0.51/0.96\\
		Circinus& 14:13:09.9 	& $-$65:20:21	& \phantom{0}4.2		& $6.9\times10^{10}$ \phantom	& AGN/SB 	&  \phantom{00}Yes\tablefootmark{(10)}	& No&3.5 & 0.37/0.81\\
		\hline
		\hline                                   
	\end{tabular}
	\tablefoot{Note.\\
		\tablefoottext{a}{Distances taken from \citet{Sanders2003} except for Circinus which value was taken from the NASA/IPAC Extragalactic Database;}
		\tablefoottext{b}{infrared luminosities taken from  \cite{Sanders2003} except for Circinus which was calculated from the flux densities of 12$\mu$m, 25$\mu$m,60$\mu$m and 100$\mu$m taken from \citet[][Table 1]{Brauher2008};}
		\tablefoottext{c}{Presence of molecular outflow report. (1) e.g. \cite{Bolatto2013,Walter2017}, (2) e.g. \cite{Emsellem2006,Barbosa2014,Garcia-Burillo2015},(3) e.g. \cite{Aalto2012a,Aalto2016},(4) e.g. \cite{Sakamoto2014,Michiyama2018},  (5) e.g. \cite{Ohyama2019}, (6) e.g. \cite{Spoon2013}}, (7) \cite{Henkel2018}, (8) e.g. \citet[][]{Aalto2019b} (9) e.g. \cite{Israel2017}, (10) \cite{Curran1999,Zschaechner2016}; 
		\tablefoottext{d}{Presence of water detection. (1) Herschel SPIRE data, (2) \cite{Gonzalez-Alfonso2012}, and Herschel SPIRE data ,(3) \cite{Gonzalez-Alfonso2010},(4) Herschel PACS data};
		\tablefoottext{e}{On source observation time};
		\tablefoottext{f}{Noise level obtained from the CLASS (http://www.iram.fr/IRAMFR/GILDAS).}
	}
\end{table*}

\noindent
Molecular gas plays an important role in the bursts of star formation and feeding of Super Massive Black Holes (SMBHs), in particular when collisions of gas-rich galaxies funnel massive amounts of material into nuclei of luminous and ultra luminous infrared galaxies (LIRGs/ULIRGs) \citep{Sanders1996,Iono2009}. 
Molecular gas probes important physical processes such as massive outflows powered by the nuclear activity \citep[e.g.][]{Sturm2011, Feruglio2011, Chung2011, Aalto2012,Aalto2012a,Cicone2014,Sakamoto2014,Aalto2015a,Fluetsch2019,Veilleux2020}. These outflows help to regulate the growth of galaxy nuclei and the implied outflow momenta suggest that the inner regions of these galaxies may be cleared of material within a few Myr \citep[e.g.][]{Feruglio2011,Cicone2014}. The morphology of the molecular feedback ranges from wide-angle winds \citep{Veilleux2013}, gas entrained by radio jets \citep{Morganti2015,Garcia-Burillo2015,Dasyra2016} to collimated molecular outflows \citep{Aalto2016,Sakamoto2017a, Barcos-Munoz2018,Falstad2019}.

The fate of the molecular gas is still not well understood. One question is whether the gas is expelled from the galaxy or it returns to fuel another cycle of activity \citep[e.g.][]{Pereira-Santaella2018,Lutz2020}. It is also not known if the molecular clouds are being swept up from a circumnuclear disk or formed in the outflow \citep[e.g.][]{Ferrara2016}. Another question is whether the clouds expand and evaporate or if they condense and form stars in the outflow \citep{Maiolino2017}. A detailed study of the physical conditions of the molecular gas in outflows is necessary to understand its origin, how it evolves within the outflow, as well as how this is linked to the nature of the driving force.

Important extragalactic probes of cloud properties include molecules such as HCN, HCO$^+$, HNC, CN and HC$_3$N. These molecules have high dipole moments which means that they normally require high densities ($n>10^4$ cm$^{-3}$) for efficient excitation. Hence they are in general more tightly related to dense star-forming gas compared to low-$J$ transitions of CO, as shown by \citet{Gao2004}. 

The detection of luminous HCN and CN emission in, often AGN-powered, molecular outflows is evidence of the presence of large amounts of dense molecular gas \citep[e.g.][]{Aalto2012,Sakamoto2014,Matsushita2015,Garcia-Burillo2015, Privon2015,Cicone2020}. (Although we should note that faint, but widespread, HCN emission may also emerge from more diffuse gas \citep[e.g.][]{Nishimura2017}.)

To understand the origin of the dense gas in outflows, we require tracers that probe specific physical and chemical conditions. One of the most reactive elements is sulphur: its chemistry is very sensitive to the thermal and kinetic properties of the gas. Sulphur-bearing molecules, such as H$_{2}$S, SO and SO$_{2}$, appear to be unusually abundant in clouds with high mass star formation, accompanied by shock waves and/or high temperatures \citep[e.g.][]{Mitchell1984,MInh1990}. Hence  observations of sulphur-bearing molecules could be used to constrain the properties of the dense gas in outflows.

H$_{2}$S is often assumed to be the most abundant sulphur-bearing species on ice mantles, as hydrogenation of sulphur should be very efficient on the dust grains, despite not being detected in interstellar ices \citep{Charnley1997, Wakelam2004, Viti2004}. In quiescent star-forming regions,  gas-phase abundances of H$_{2}$S are therefore low because they are depleted on grains.

Strong shocks can sputter H$_{2}$S off the grains moving into the gas phase. There are many theoretical studies to test this hypothesis \citep[e.g.][]{Woods2015}, giving ample evidence of enhanced gas phase H$_{2}$S in warm and/or shocked environments. In model calculations,  \citet{Holdship2017} have shown that as a C-type shock passes through the cloud, sulphur is first transformed into H$_{2}$S and its abundance increases greatly in the post-shock gas. H$_{2}$S chemistry and grain-processing features share several properties with that of water, H$_{2}$O, but with the added advantage that its ground state transition occurs at a wavelength of $\lambda$=2~mm and can be observed with ground-based telescopes.

H$_{2}$S seems to be a  promising tracer to probe outflows and nuclear, dusty activity in galaxies. Photon Dominated Regions (PDRs) may also contribute significantly to the H$_{2}$S emission \citep[e.g.][]{Goicoechea2021}. Here we present the results of an Atacama Pathfinder Experiment (APEX) single dish telescope pilot survey of the 168.7 GHz $1_{1,0}$--$1_{0,1}$ transition of H$_{2}$S toward a sample of galaxies to study the relationship between H$_{2}$S emission and the outflow activity of the galaxies. We also present 168.7 GHz IRAM Northern Extended Millimiter Array (NOEMA) H$_{2}$S observations of the nucleus and outflow of the ULIRG merger Mrk~231.

The paper is organised as follows. In Sec.~\ref{s:observations} we present the galaxy sample, as well as the APEX and NOEMA observation, and in Sec.~\ref{s:results} the results. In Sec.~\ref{s:discussion} we discuss the luminosity of the H$_{2}$S emission in relation to that of other tracers such as HCN, HNC, HCO$^+$, H$_2$O and CH$_3$CN and also in relation to the presence of an outflow. In Sec.~\ref{s:conclusions} we present our conclusions and an outlook.



\section{Observations}
\label{s:observations}

\subsection{The sample}
\label{subsec:sample}

In Table~\ref{tab:Basic} the sources, consisting of a total of 13 (12 are observed with APEX and one with NOEMA) galaxies, are presented. They were selected as a mixture of starburst- and AGN-powered luminous infrared galaxies, which are observable with APEX. Some have detected molecular outflows while other have no indications of outflows (Table~\ref{tab:Basic}). The objects are nearby ($D<60$ Mpc) to allow us to detect H$_{2}$S emission with a single dish telescope. The Mrk~231 observations stem from a separate NOEMA observational program that is included in this paper since it fits with the observational goals of the APEX program. Mrk~231 is more distant than the objects observed with APEX.

\subsection{APEX SEPIA B5}
\label{subsec:obs_apex}

The observations were carried out with APEX SEPIA Band 5 under program codes 096.F-9331(A), 099.F-9312(A) and 0100.F-9311(A). NGC~4945, Circinus, NGC~3256, NGC~1377 and IC~860 were observed in September and November 2015. NGC~1068, NGC~253, NGC~4418, M~83, NGC~4826, NGC~5128 and NGC~1097 were observed in June and August 2017. The basic information of the sources and the observational details are listed in Table \ref{tab:Basic} and Table \ref{tab:lines}, respectively. The heterodyne SEPIA band 5 receiver, which covers the frequency range from 159 to 211~GHz, was used as a frontend. In the backend we used the PI-XFFTS spectrometer, which consists of two units that provide a bandwidth of 4 GHz and 32,768 spectral channels each. The observations were done in the position-switching mode in the frequency ranges 165-169 GHz and 177-181 GHz which allowed for a simultaneous search for emission lines of e.g., H$_{2}$S, HCN and HCO$^{+}$. The typical precipitable water vapor level during the observations was $\sim$1-2 mm. The temperature scale of the data is  in units of the antenna temperature corrected for the atmospheric attenuation, $T^{\prime}_{\mathrm A}$. The full width at half maximum (FWHM) of the sidebands of the telescope can be computed approximately as $FWHM_{\rm{LSB}}$=$37\rlap{.}^{\prime\prime}4\times(167 / [\nu_{\rm obs} /\rm{GHz}])$ and $FWHM_{\rm{USB}}$=$34\rlap{.}^{\prime\prime}9\times(179 / [\nu_{\rm obs} /\rm{GHz}])$, respectively, where $\nu_{\rm obs}$ is the observed frequency. 

CLASS\footnote{See \url{http://www.iram.fr/IRAMFR/GILDAS} for more information about the GILDAS software.}  was used for further analysis including baseline subtraction of a polynomial of order 1 for individual scans and spectral smoothing. The native channel widths of the observations ranged from 0.07 km s$^{-1}$ for USB to 0.14 km s$^{-1}$ for LSB. The data were smoothed to a final velocity resolution of 50 km~s$^{-1}$.

\def\degr{\hbox{$^\circ$}}
\def\arcmin{\hbox{$^\prime$}}
\def\arcsec{\hbox{$^{\prime\prime}$}}

\begin{table}
	\caption{List of molecular lines in the APEX band\label{tab:lines}}
	\begin{center}
		\begin{tabular}{lccc}
			\hline
			Molecule & Transition & Rest Frequency & Sideband\\
			&&(GHz)&\\
			\hline
			\\
			CH$_3$OH &	$1_1$--$1_0$ 		&165.050 &LSB\\
			CH$_3$CN & 9--8				&165.569 &LSB\\
			H$_{2}$S & $1_{1,0}$--$1_{0,1}$	&168.763 &LSB\\
			H$_2$CS & 5--4 				&169.113 &LSB\\
			HCN & 2--1					&177.261 &USB\\
			HCO$^{+}$ & 2--1				&178.375 &USB\\
			SO & $10_{10}$--$11_{10}$		&178.615 &USB \\
			HOC$^{+}$ & 2--1				&178.972 &USB\\
			HNC & 2--1					&181.324 &USB\\
			\hline
		\end{tabular}
	\end{center}
\end{table}%

\subsection{NOEMA observations}
\label{subsec:obs_noema}

Observations of \object{Mrk~231} were carried out with the NOEMA on April 21, 2017. Data were obtained with eight antennas in the most compact configuration, with baselines between 24 and 176~m. The phase centre of the observations was located at $\alpha$=12:56:14.2 and $\delta$=+56:52:25. The 2~mm-band receivers were tuned to 161.934~GHz to cover the H$_{\rm 2}$S\,1$_{\rm 10}-1_{\rm 01}$ line in the 3.6~GHz bandwidth of WideX. The instrumental spectral resolution was 1.95~MHz ($\sim$3.6~km\,s$^{\rm -1}$). For analysis purposes we smoothed the data to $\sim$20~MHz channel spacing ($\sim$37.0~km\,s$^{\rm -1}$), resulting in 1$\sigma$ rms noise levels per channel of $\sim$0.6~mJy\,beam$^{\rm -1}$. During the observations, different sources were observed for calibration purposes: \object{MWC~349} as flux calibrator, \object{3C~454.3} as bandpass calibrator, and \object{J1418+546} and \object{J1300+580} as phase calibrators.\\
\indent
Data reduction and analysis were performed using the CLIC and MAPPING software packages within  GILDAS\footnote{\url{http://www.iram.fr/IRAMFR/GILDAS}}, and AIPS\footnote{\url{http://www.aips.nrao.edu/index.shtml}}. Applying a natural weighting scheme led to a beam size of 2.91\arcsec\,$\times$\,2.13\arcsec, with position angle 75.64\degr.



\section{Results} 
\label{s:results}

\subsection{APEX}
\label{subsec:res_apex}

Several molecular lines are detected towards the targeted galaxies (see Table~\ref{tab:lines}). Of the 12 galaxies in the APEX sample, 11 were detected in HCN and HCO$^{+}$ 2--1, and nine galaxies were detected in the target  H$_{2}$S line. In addition, we also detected eight galaxies in SO, seven in HNC  and CH$_{3}$OH, four in CH$_{3}$CN, H$_{2}$CS and HOC$^{+}$. Spectra are shown in Fig. \ref{append1}. The integrated line intensities are presented in Table \ref{tab:APEXresult}.

\begin{table}
	\caption{List of molecular lines in the NOEMA band (Mrk~231)\label{tab:NOEMA_lines}}
	\begin{center}
		\begin{tabular}{lcc}
			\hline
			Molecule & Transition & Rest Frequency \\
			&&(GHz)\\
			\hline
			\\
			HC$_5$N &	63--62		 			&167.718 \\
			CH$_3$OH$^{\dagger}$ & $9_1$--$9_0$	&167.931 \\
			CH$_3$OH & 	$v_2$$=$$2$, $10_8^+$-- $11_5^+$	&168.076 \\
			CH$_3$OH$^{\dagger}$ & 	$40_3$--$39_5$	&168.296 \\
			C$_2$S & $13_{13}$--$12_{12}$ 					&168.407 \\
			H$_{2}$S & $1_{1,0}$--$1_{0,1}$		&168.763 \\
			H$_2$CS & 5--4 				&169.113 \\
			CH$_3$OH & 	$10_1$--$10_0$				&169.335 \\
			CH$_3$OH & 	$15_1$--$14_3$				&169.488 \\
			C$_2$S$^{\diamond}$ & $13_{14}$--$12_{13}$ 		&169.753 \\
			CH$_3$OH & 	$3_2$--$2_1$				&170.060 \\
			HC$^{18}$O$^{+}$ & 2--1			&170.323 \\
			CH$_3$CCH & 10--9				&170.876 \\
			\hline
		\end{tabular}
		\tablefoot{$^{\dagger}$ A possible alternative identification here is H$_2^{34}$S.$^{\diamond}$Alternative identifications are NH$_2$CN or $^{34}$SOH. }
	\end{center}
\end{table}

\begin{table*}
	\centering
	\caption{NOEMA observation result for Mrk~231 \label{tab:NOEMAresult}}
	\begin{center}
		\begin{tabular}{llccccc}
			\hline\hline
			&			&\multicolumn{3}{c}{Core}													&\multicolumn{2}{c}{Wings}\\
			Species	&Transition	&Line peak ($T_{\mathrm{A,peak}}$)	&$I$\tablefootmark{a}		& $\Delta v$ &$I_{\rm{redwing}}$\tablefootmark{b}	&$I_{\rm{bluewing}}$\tablefootmark{c}\\
			&					&(mJy beam$^{-1}$)			&(Jy kms$^{-1}$)			&(kms$^{-1}$)	&(Jy kms$^{-1}$)				&(Jy kms$^{-1}$)\\
			\hline
			\\
			H$_{2}$S	&$1_{1,0}$--$1_{0,1}$&7.08					&1.69$\pm$0.11				&224.68$\pm$29.48	&0.07$\pm$0.07				&0.02$\pm$0.07\\
			HCN		&$J$=2--1		&90.6					&24.09$\pm$0.14				&255			&1.74$\pm$0.14					&1.18$\pm$0.14\\
			HCO$^{+}$&$J$=2--1		&68.0					&15.58$\pm$0.14				&215			&0.78$\pm$0.14					&1.09$\pm$0.14\\
			\hline
		\end{tabular}
	\end{center}
	\tablefoot{The values for HCN and HCO$^{+}$ are taken from \cite{Lindberg2016}. (a) Integrated intensity between -250 and 250 km s$^{-1}$. 
		(b) Integrated intensity between 350 and 515 km s$^{-1}$ for H$_{2}$S, between 350 and 990 km s$^{-1}$ for HCN and HCO$^{+}$. 
		(c) Integrated intensity between -538 and -350 km s$^{-1}$ for H$_{2}$S, between -990 and-350 km s$^{-1}$ for HCN and HCO$^{+}$.}
\end{table*}

\begin{table*}
	\centering
	\caption{Integrated intensity ratios}
	\label{tab:ratio}
	\begin{tabular}{l c c c c c c}
		\hline\hline
		Source & $R_\frac{\rm{HCN}}{\rm{HCO^{+}}}$ & $R_\frac{\rm{H_{2}S}}{\rm{HCN}}$   & $R_\frac{\rm{H_{2}S}}{\rm{HCO^{+}}}$ & $R_\frac{\rm{CH_{3}OH}}{\rm{HCN}}$ & $R_\frac{\rm{CH_{3}CN}}{\rm{HCN}}$ & $R_\frac{\rm{SO}}{\rm{HCN}}$\\
		\hline
		\multicolumn{7}{c}{from APEX observations}\\
		\hline
		\\
		NGC~3256 	& 0.54$\pm$0.11 & 0.12$\pm$0.03 & 0.07$\pm$0.01 & >0.03$\pm$0.01 & <0.01$\pm$0.01 & 0.06$\pm$0.02 \\
		NGC~1068	& 1.39$\pm$0.30 & 0.05$\pm$0.01 & 0.08$\pm$0.02 & >0.02$\pm$0.01 &  <0.02$\pm$0.01 & 0.02$\pm$0.01 \\
		IC~860 		& >6.31$\pm$6.39 & <0.10$\pm$0.11 & <0.62$0.93\pm$ & <0.07$\pm$0.09 & 0.48$\pm$0.13 & <0.16$\pm$0.16 \\
		NGC~4418	& 1.73$\pm$0.37 & 0.17$\pm$0.08 & 0.30$\pm$0.14 & ---& 0.19$\pm$0.08 & 0.12$\pm$0.05 \\
		Circinus 	& 0.86$\pm$0.18 & 0.13$\pm$0.03 & 0.11$\pm$0.02 &  >0.03$\pm$0.01 & <0.01$\pm$0.01 & 0.03$\pm$0.01 \\
		NGC~1097 	& 1.21$\pm$0.26 & 0.05$\pm$0.02 & 0.06$\pm$0.02 &  >0.02$\pm$0.02 & <0.02$\pm$0.02 & <0.02$\pm$0.02 \\
		NGC~4945	& 1.04$\pm$0.22 & 0.05$\pm$0.01 & 0.05$\pm$0.01 & >0.06$\pm$0.01 & 0.03$\pm$0.01 & <0.01$\pm$0.01 \\
		NGC~253	& 0.99$\pm$0.21 & 0.09$\pm$0.02 & 0.09$\pm$0.02 &  >0.06$\pm$0.01 & 0.03$\pm$0.01 & 0.03$\pm$0.01 \\
		M~83 		& 0.94$\pm$0.20 & 0.08$\pm$0.02 & 0.07$\pm$0.02 &  >0.07$\pm$0.02 & <0.02$\pm$0.01 & <0.03$\pm$0.03\\
		NGC~4826 	& 1.12$\pm$0.24 & 0.18$\pm$0.04 & 0.20$\pm$0.04 &  <0.03$\pm$0.03 & <0.03$\pm$0.03 & <0.06$\pm$0.07\\
		\hline
		\multicolumn{7}{c}{from NOEMA observations}\\
		\hline
		Mrk~231 (Core) & 1.55 & 0.07 & 0.11 & ---&---&---\\
		\phantom{Mrk~231} (B.wing)	& 1.08 & 0.02 & 0.02 & ---&---&---\\
		\phantom{Mrk~231} (R.wing)	& 2.23 & 0.04 & 0.09 & ---&---&---\\
		\hline\hline
	\end{tabular}
	\tablefoot{The line intensity ratio of two molecular lines, $R_{line1/line2}$ from APEX and NOEMA observations. Transitions of the lines are: HCN 2-1, HCO$^{+}$ 2-1, H$_{2}$S ${1_{1,0}-1_{0,1}}$, CH$_{3}$OH ${1_{1}-1_{0}}$, CH$_{3}$CN 9-8, and SO ${10_{10}-11_{10}}$. The values for the blue and red wings for Mrk~231 are calculated using the integrated intensities between the same velocity ranges as in Table~\ref{tab:NOEMAresult}.}
	
\end{table*}

\subsection{NOEMA - Mrk~231}
\label{s:M231}

\begin{figure}
	\begin{center}
		\includegraphics[angle=0,scale=0.38]{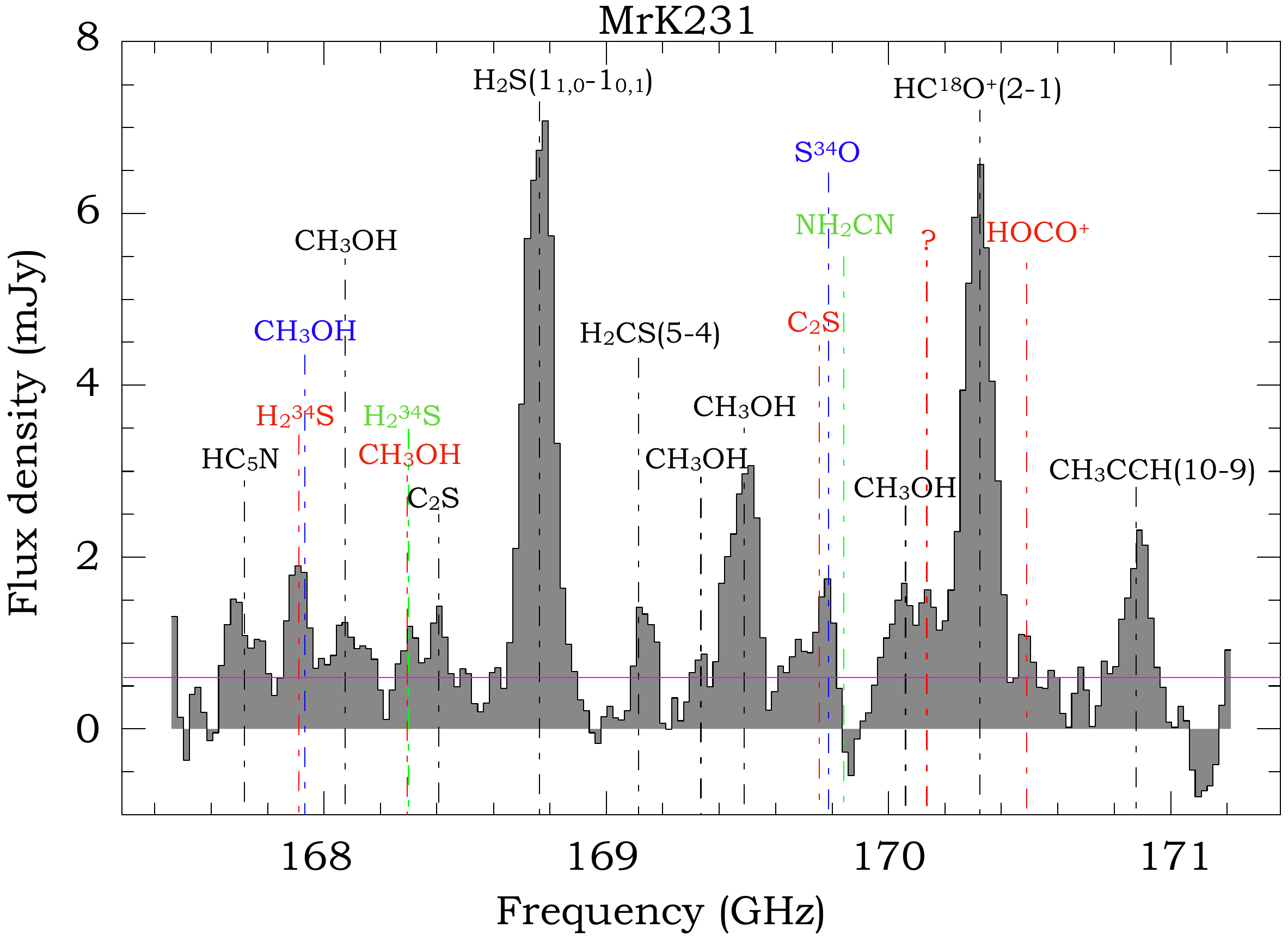}
		\caption{NOEMA broad-band (3.6 GHz), 2mm spectrum of Mrk~231, centred on H$_{2}$S. The x-axis is rest frequency and the y-axis is flux in 
			mJy. Identified lines have detection thresholds above 3$\sigma$ (rms, indicated in the magenta line).}\label{fig:full231}
	\end{center}
\end{figure}

\begin{figure}
	\begin{center}
		\includegraphics[angle=0,scale=0.38]{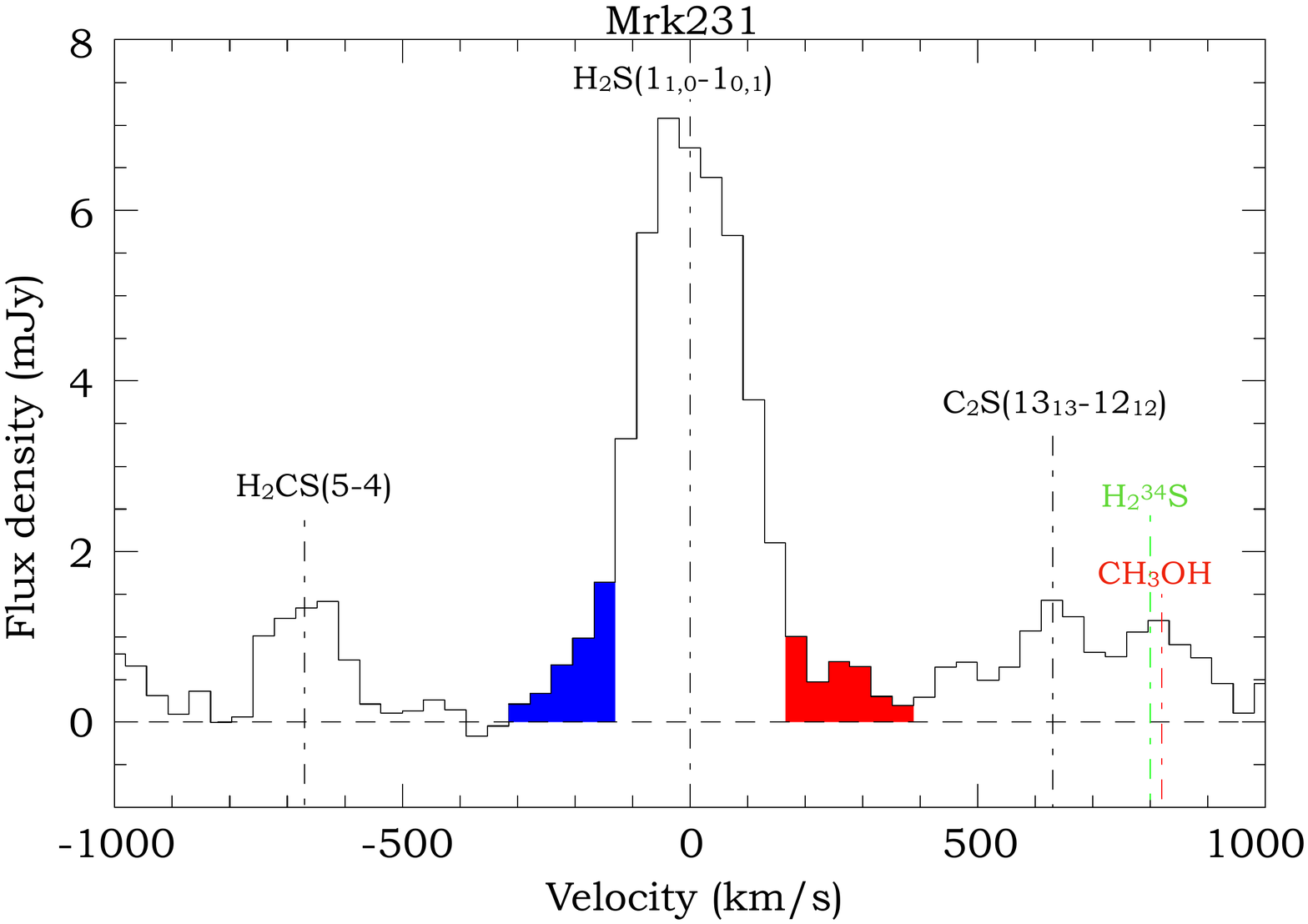}
		\caption{NOEMA H$_{2}$S spectrum of Mrk~231 (zoomed in from Fig.\ref{fig:full231}). The x-axis is in km/s relative to the central velocity 
			(12642 km/s).
			Red- and blue-shifted line-wing emission is indicated with corresponding colours. Other identified lines are indicated in the figure 
			(same as in Fig.\ref{fig:full231})}\label{fig:H2S231}
	\end{center}
\end{figure}

Fig. \ref{fig:full231} shows the full bandwidth NOEMA spectrum of Mrk~231. We have labelled all molecular lines that are brighter than 3$\sigma$ above noise level and they are also listed in Table~\ref{tab:NOEMA_lines}. H$_{2}$S is clearly detected with a peak intensity of $\sim$7~mJy. The HC$^{18}$O$^{+}$ 2--1 line is also bright with a peak intensity almost comparable to that of H$_{2}$S. H$_{2}$CS 5-4, which is also detected in NGC~4945, NGC~4418, NGC~253 and NGC~1068 with APEX, shows a tentative detection. In Fig. \ref{fig:H2S231} we show a spectrum zoomed in on the H$_{2}$S line in Mrk~231. It shows that  H$_{2}$S is detected in the line wings up to velocities of -300 km s$^{-1}$ on the blueshifted side and possibly out to 500 km s$^{-1}$ on the redshifted side. The integrated line intensity of H$_{2}$S is presented in Table\ref{tab:NOEMAresult}.

\section{Discussion}
\label{s:discussion}

This is the first survey of the $\lambda$=2~mm $1_{10}$--$1_{01}$ line of o-H$_{2}$S in luminous infrared galaxies. In this paper we report eight new detections, seven with APEX and one from NOEMA observations. To date, only three detections of 2~mm H$_{2}$S line emission have been reported outside our own galaxy : Large Magellanic Cloud (LMC) \citep[e.g.][]{Heikkila1999},  NGC~253 \citep[e.g.][]{Martin2005} and Arp~220 \citep[e.g.][]{Martin2011}. Our sample, which is chosen from nearby LIRGs (see Sec.~\ref{subsec:sample}), includes a variety of galaxies in terms of morphology and kinematics.

\subsection{H$_{2}$S formation mechanisms}
\label{subsec:formation}

H$_{2}$S can be formed through hydrogenation of a sulphur atom on grain surfaces, which can then be either i) thermally desorbed by shocks \citep[][Fig. 2]{Esplugues2014} or by IR radiation \citep[e.g.][]{Crockett2014}, ii) photodesorbed by UV radiation, iii) desorbed after cosmic ray impacts, and iv) sputtered by shocks. Chemical desorption could be also efficient to release H$_{2}$S into the gas phase \citep[e.g.][]{Oba2019,Navarro-Almaida2020}, but it is not a dominant mechanism when we focus on starburst and AGN-host galaxies. Hydrogenation of sulphur is also possible in the gas-phase if the kinetic temperature is higher than a few to several 1000 K \citep{Mitchell1984}.
For gas densities in excess of $n$=$10^4$ cm$^{-3}$, gas and dust are coupled and thermalised with similar temperatures. Therefore, under the assumption that most of the H$_{2}$S emission is emerging from dense gas, which kinetic temperature can be between 10 to 300 K, hydrogenation in the gas-phase is unlikely because it is unlikely that the bulk of the dense gas can be maintained at such a high temperature. On smaller scales, close to heating sources, such high temperatures are however possible.

\subsubsection{ H$_{2}$S enhancement via shocks} 
\label{s:shocks}

\begin{figure}
	\begin{center}
		\includegraphics[angle=0,width=0.43\textwidth]{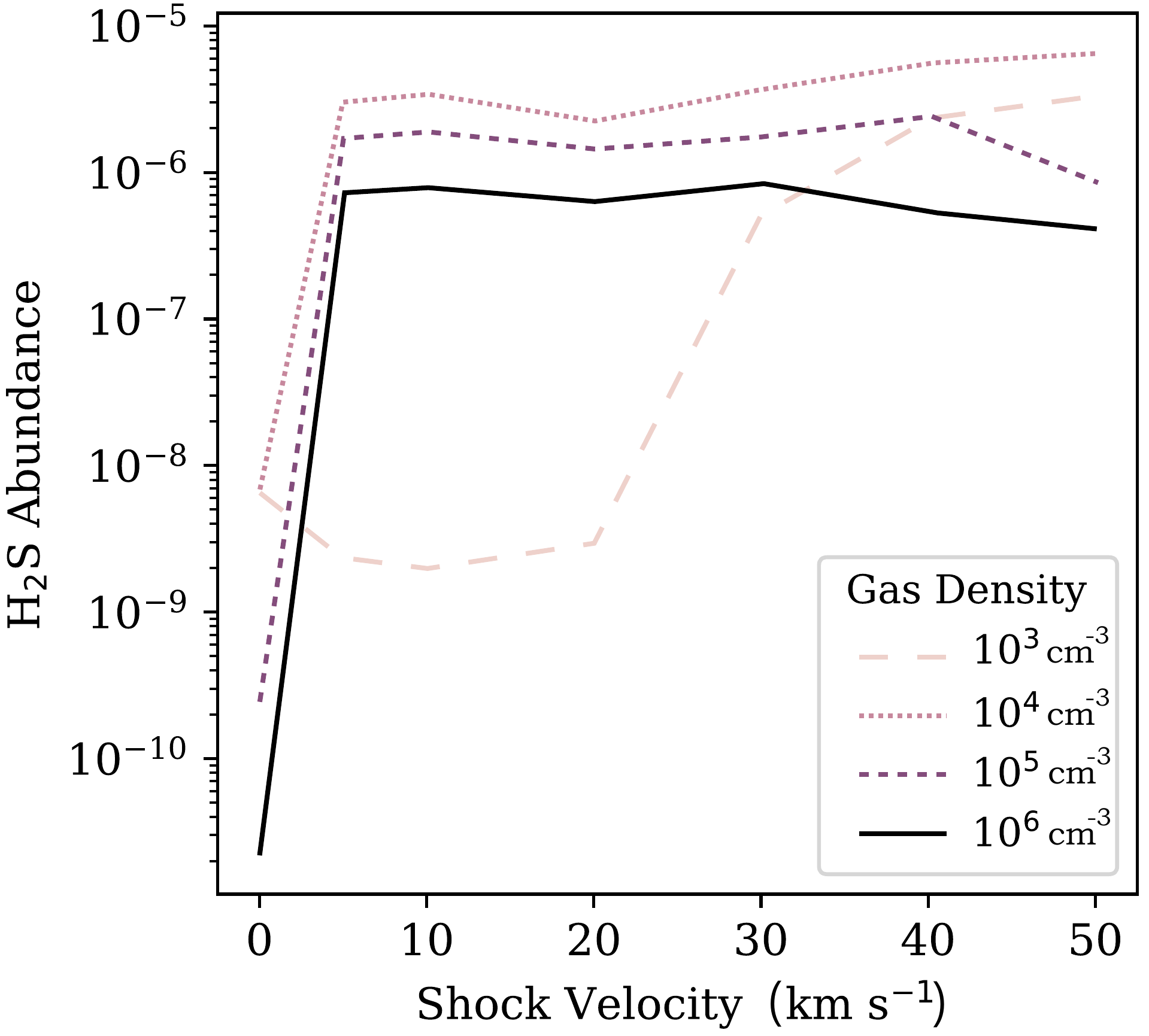}
		\caption{Average fractional abundance as a function of shock velocity for H$_{2}$S from \cite{Holdship2017}. Each line shows a different pre-shock density; the log($n_{\rm H}$) values are
		given in the lower right plot.}\label{fig:H2Sshock}
	\end{center}
\end{figure}

Shocks can cause both thermal desorption and mantle sputtering. For the case where the sulphur species are sputtered into the gas-phase by strong shocks, Fig.\ref{fig:H2Sshock} shows the average fractional abundance as a function of shock velocity for H$_{2}$S \citep{Holdship2017}. In this figure, it can be seen that for low pre-shock gas densities (10$^{3}$ cm$^{-3}$), the fractional abundance is 10$^{-8}$ for shock velocities less than 20 km s$^{-1}$, and increases by a factor of $\sim$100 for shock velocities $>$30 km s$^{-1}$. For higher pre-shock gas densities ($>$10$^{4}$ cm$^{-3}$), the abundance reaches values  $\sim$10$^{-6}$ even for shock velocities as low as $\sim$5 km s$^{-1}$. This suggests that a high abundance of this molecule could be due to the passage of a shock in the absence of other mechanisms capable of releasing H$_{2}$S from the ices. The caveat here is that this model considers only the case when most of the sulphur hydrogenates on the grains.

According to chemical models, H$_{2}$S transforms quickly into SO, SO$_2$, etc., within a timescale of $\leq 10^{4}$ yr \citep[e.g. ][]{PineaudesForets1993}. Shock elevation of H$_{2}$S abundances therefore requires continuous replenishment, for example by internal shocks in the outflow.  Internal shocks can occur if the molecular gas is interacting with hotter material in the outflow or if parts of the outflow is interacting with slower/faster moving gas components, for example due to periodic ejection processes. It is also possible that H$_{2}$S is enhanced in highly turbulent gas. If so, the link to the outflow activity becomes less clear and may depend on whether the outflow process is feeding turbulence in the gas in a fountain scenario \citep[e.g.][]{Aalto2020}. Other sources that drive turbulence include accretion and inflow, galactic rotation and stellar feedback \citep{Klessen2016}.

\subsubsection{H$_{2}$S enhancement via thermal desorption, uv radiation and cosmic rays}
\label{s:radiation}

As discussed above, other possible method for getting H$_{2}$S off the grains and into the gas-phase is by thermal desorption from warm grains, non-thermal UV photo-desorption, and cosmic-ray-induced desorption.  In a recent Herschel study of higher rotationally excited H$_{2}$S lines in Orion hot core, \citet{Crockett2014} pointed to an H$_{2}$S origin in dense, embedded gas, in close proximity to a hidden self-luminous protostellar source (i.e. heating the grains above the H$_{2}$S sublimation temperature). In addition, \citet{Goicoechea2021} discussed and modelled the photodesorption of H$_{2}$S in UV-illuminated environment (PDRs) and this is a possible additional H$_{2}$S formation mechanism for some of the sample galaxies. However, for some objects the gas and dust column densities will be too high to allow the UV radiation to penetrate very far.  A further possible mechanism is cosmic-ray-induced desorption. These scenarios are further discussed in Sec.~\ref{s:HotCore}.

\subsection{Column density and abundance}
\label{s:column}

In this section we make simple estimates of column densities and abundances for basic comparisons with other sources and shock models. We assume collisional excitation for the H$_{2}$S molecule throughout the paper \citep[e.g.][]{Mangum2015}\footnote{$n_{\rm crit,H_2S}\approx7.8\times 10^5 [\rm cm^{-3}]$,  $n_{\rm crit,HCN(2-1)}\approx1.3\times 10^7 [\rm cm^{-3}]$, $n_{\rm crit,HCO^{+}(2-1)}\approx1.2\times 10^6 [\rm cm^{-3}]$ and $n_{\rm crit,CO(1-0)}\approx2.2\times 10^3 [\rm cm^{-3}]$ at $T_{\rm k}$=30 K.See Appendix \ref{append2}.}. H$_{2}$S may also be pumped by infrared radiation, but \citet{Crockett2014} argued that IR pumping for the transitions of H$_{2}$S lower than $J$=3 is unlikely because these transitions occur at longer wavelengths ($\lambda > 100 \mu m$).

Column densities of H$_{2}$S, HCN and HCO$^{+}$ molecules for each galaxy were calculated using the formalism in \citet{Mangum2015, Mangum2016, Mangum2017}. Since we have no direct information on the excitation of the H$_{2}$S molecule nor the distribution of its emission, we have to make some assumptions to estimate the column densities. We have to assume an excitation temperature, $T_{\rm ex}$, and we estimated column densities for three different values of $T_{\rm ex}$ of 30 K, 60 K and 150 K to compare the results. This range of temperature is reflecting the span of kinetic temperature typically found in the inner regions of galaxies \citep[e.g.][]{Aalto1995, Mangum2013a, Mangum2019, Salak2014}. This is often an upper limit to the excitation temperature. We also assumed a beam-filling factor of unity f$_{\rm s}$=1 and optically thin ($\tau<<1$) emission for all lines for simplicity.  We compare the H$_{2}$S to the HCN abundance, assuming that both lines emerge from the same dense gas component and have the same $T_{\rm ex}$ (see Sec.\ref{subsec:caveats} for a discussion of caveats). The effect of these simplifying assumptions are to some extent cancelled out when we use the same assumption for all species and compare their ratios. For example, we find that the difference between resulting column density ratios for different $T_{\rm ex}$ is small, less than a factor of 2.

We consider the two most extreme cases in the observed H$_{2}$S/HCN line ratios: NGC~4418 and NGC~1097 (see Table~\ref{tab:ratio}).  We estimate the H$_{2}$S/HCN abundance ratio (=column density ratio) to be $X$(H$_{2}$S/HCN)$\sim$0.1 for NGC~4418 and $\sim$0.03 for NGC~1097 (for $T_{\rm ex}$=60 K). We choose the column density values derived for an $T_{\rm ex}$ of 60 K since it is an intermediate value for $T_{\rm ex}$. As note above, the impact of the $T_{\rm ex}$ on the column density ratio is quite small.

The difference in abundance ratio reflects quite well the difference in line intensity ratio between the two galaxies. We will therefore use the integrated intensity of H$_{2}$S relative to HCN and HCO$^{+}$ in Sections \ref{s:shocktest} and \ref{s:HotCore} to indicate potential H$_{2}$S enhancements in relation to the dense gas.  The estimated values of $X$(H$_{2}$S/HCN) seem reasonable compared to those found for the Galactic Orion star forming complex.  The value for  $X$(H$_{2}$S/HCN) varies strongly within the region, from $<10^{-4}$ in the ridge (quiescent gas) to $\sim$0.3 for the plateau (shocked gas) \citep{Blake1987} while the $X$(H$_{2}$S/HCN) for the Orion hot core may be greater than unity.

We also determined the H$_{2}$S abundance in relation to H$_2$ to $X$(H$_{2}$S)$\approx1.3\times10^{-7}$ and $X$(H$_{2}$S)$\approx1.7\times10^{-8}$ for NGC~4418 and NGC~1097, respectively. To obtain a value for  $N$(H$_2$), we obtained the molecular mass from the CO 1--0 luminosity\footnote{CO 1--0 luminosities taken from \citet{Sanders1991} for NGC~4418 and \citet{Gerin1988} for NGC~1097.} assuming a standard CO-to-H$_2$ conversion factor \citep{Bolatto2013a}. Assuming that the molecular mass was uniformly distributed in the APEX beam for our H$_{2}$S observations we could calculate an average $N$(H$_2$). Comparing to the shock model results presented in Fig.\ref{fig:H2Sshock} shows that our simple estimation for $X$(H$_{2}$S) is consistent with a shock origin of the observed H$_{2}$S for NGC~4418. On the other hand, \citet{Crockett2014} show that these enhanced H$_2$S abundances are caused by strong dust heating and subsequent thermal desorption in hot cores. In Sections \ref{s:shocktest} and \ref{s:HotCore} we compare line ratios with dynamical and radiative properties of the galaxies to see if we can determine if outflows may be the cause behind H$_{2}$S enhancements in LIRGs.
	
\subsubsection{Caveats}
\label{subsec:caveats}

Our column density values are strongly affected by the assumptions on beam-filling factors and $T_{\rm ex}$. They give average values at best which is acceptable in searching for trends and correlations, but {\it further high resolution and multi-transition follow-up studies are required to chart the underlying causes behind H$_{2}$S intensity enhancements.} One example to illustrate this is NGC~4418. \citet{Costagliola2015} studied the molecular excitation temperatures and column densities of many species in NGC~4418 with high resolution and multi-transition observations using ALMA. They could compensate for the high optical depth of the HCN lines with a population diagram analysis, and (with an optically thin assumption for H$_{2}$S) estimated $X$(H$_{2}$S/HCN)$\sim$0.5. Should we use their LTE values, uncorrected for opacity, then we would arrive at $X$(H$_{2}$S/HCN)$>60$. The discrepancy of this latter value compared to our result is that \citet{Costagliola2015} use a $T_{\rm ex}$ of 7 K for HCN and 70 K for H$_{2}$S while we use the same value for both lines. HCN may exist also in more extended and colder region compared to H$_{2}$S, which contribute to lowering the excitation.

\subsection{Testing the notion of shock-enhanced H$_{2}$S}
\label{s:shocktest}
Out of the 13 sample LIRGs, 10 have been reported to have molecular outflows (See Table\ref{tab:Basic}). In Sec.~\ref{s:shocks} we discussed a possible link between H$_{2}$S enhancement and shocks in galaxies, for example caused by large scale outflows. The sample galaxies display a range in outflow velocity, from slow (150 km s$^{-1}$) to fast ($>$250 km s$^{-1}$).

To search for evidence of elevated H$_{2}$S emission related to shock events, we put together line intensity ratios of H$_{2}$S/HCO$^{+}$ and H$_{2}$S/HCN from our observation by APEX and NOEMA (see Table~\ref{tab:ratio})(see discussion in Sec.~\ref{s:column}).

\subsubsection{APEX single dish data}
\label{s:outflow1}

We identify four galaxies with particularly elevated (a factor of two to three higher than in the rest of the sample) H$_{2}$S/HCN ratios: Circinus, NGC~3256, NGC~4826 and NGC~4418. Of these, NGC~3256 has a fast molecular outflow ($>$250 km s$^{-1}$), Circinus and NGC~4418 have slow outflows ($\approx$150 km s$^{-1}$) while NGC~4826 has no report of a molecular outflow. Another example is the Seyfert galaxy NGC~1068, which has a striking AGN-driven molecular outflow \citep[e.g.][]{Garcia-Burillo2014}, but has a lower H$_{2}$S/HCN ratio compared to NGC~4826, or M83 that lack known molecular outflows. In addition, both NGC~4945 and NGC~253 have slow molecular outflows with velocities of 50~km~s$^{-1}$, but their H$_{2}$S line luminosity relative to HCN is quite different from each other (Table~\ref{tab:ratio}). 

We find that the line ratios show considerable variation among our sample galaxies, but there is no apparent indication that H$_{2}$S is more abundant in the galaxies with outflows, nor can we find any link to the outflow velocity. The lack of a clear link between the H$_{2}$S/HCO$^{+}$ and H$_{2}$S/HCN line ratios and the presence, and properties, of an outflow could be because the link is not there, or because the large APEX beam (multi-kpc scale) includes too much extended gas that is unconnected to the outflow. Indeed, H$_{2}$S absorption has been detected even in diffuse clouds \citep{Lucas2002} which may also contribute and dilute our results. Furthermore, the single dish data lack the sensitivity and stability necessary to compare the relative H$_{2}$S enhancement in the line core to that in the line wings.

Therefore it is important to investigate potential enhancements of H$_{2}$S in outflows with high resolution observations, both spectroscopically and spatially, for a sample of galaxies.


\begin{figure*}
    \begin{center}
	\includegraphics[angle=0,scale=1]{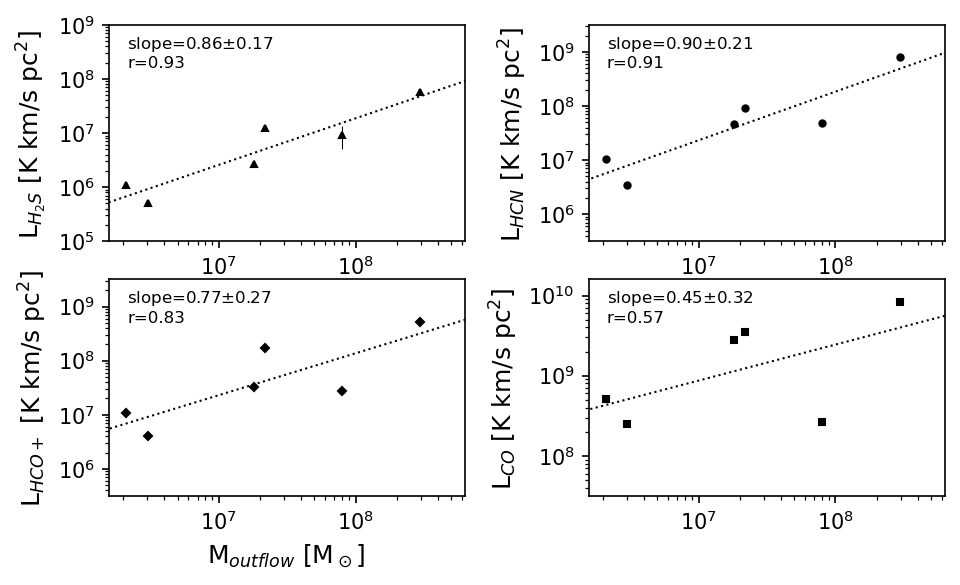}
	\caption{Line luminosities of H$_{2}$S (top left), HCN (top right) and HCO$^{+}$ (lower left) 2--1, and CO 1--0 (lower right), of Circinus, NGC~3256, NGC~1068, NGC~253, NGC~4418 and Mrk~231 are plotted against the molecular mass of the outflow, $M_{\rm outflow}$, taken from \citet{Fluetsch2019}. The correlation coefficient between the line luminosity and $M_{\rm out}$ is given in the upper left corner of each panel. Here, also the slope of the line is given.}\label{fig:L_Mof} 
    \end{center}
\end{figure*}

\subsubsection{Mrk~231 with NOEMA}
\label{s:outflow2}

To search for indications of shock-chemistry in an outflow with higher spatial resolution, we observed the 168 GHz ground state H$_{2}$S line with the NOEMA telescope toward Mrk~231.

The ULIRG quasar Mrk~231 has an extremely strong and fast molecular outflow where CO has been detected out to velocities of  750 km~s$^{-1}$ \citep{feruglio2010}. The molecular outflow also shows remarkably luminous HCN emission which is suggested to be caused by large masses of dense molecular gas, but also by elevated HCN abundances \citep{Aalto2012,Aalto2015}. It is suggested that the HCN enhancement is caused by the molecular gas being compressed and fragmented by shocks in the outflow. 

We detected the H$_{2}$S line with a peak intensity of $\approx$7.2~mJy  (Fig. \ref{fig:H2S231}) and the line has wings at both sides indicating H$_{2}$S emission in the outflow at least out to velocities of -300 kms$^{-1}$ and +500 kms$^{-1}$ (see Sec.~\ref{s:M231}). However, when we compare the relative intensity of the H$_{2}$S emission to that of HCN 2--1, we see no sign of an elevated H$_{2}$S intensity in the line wings with respect to the core. Therefore, the relatively faint H$_{2}$S emission appears inconsistent with the notion of dominant shock-chemistry in the molecular outflow of Mrk~231. More studies of the molecular formation and destruction processes of the Mrk~231 are necessary (See for example the recent detection of CN in the outflow by \citet{Cicone2020}).

It is interesting to compare Mrk~231 to the H$_{2}$S-luminous merger LIRG NGC~3256. This galaxy exhibits two strong and fast outflows: one from each nucleus \citep[e.g.][]{Sakamoto2014}. NGC~3256 has a distinct H$_{2}$S enhancement in relation to HCN and we require high resolution observations to determine if the enhancement is linked to any, or both, of the two outflows, or to some other part of its molecular emission.

\subsection{H$_{2}$S enhancement caused by thermal desorption: The "Hot Core"-like nuclear obscuration of NGC~4418} 
\label{s:HotCore}

H$_{2}$S enhancements may also be caused by other processes such as thermal desorption (see Sec.~\ref{s:radiation}). For example, in a high resolution study of H$_{2}$S towards NGC~253,  \citet{Minh2007} argue that H$_{2}$S enhancements might be associated with young massive star formation, which may imply a scenario of radiative thermal desorption.

NGC~4418 and IC~860 are suggested to host so called Compact Obscured Nuclei (CONs) in their central regions \citep[e.g.][]{Costagliola2011, Costagliola2013, Sakamoto2013, Falstad2019, Aalto2019b}. A CON is a region of exceptionally high H$_2$ column density ($N$(H$_2$)$>10^{25}$ cm$^{-2}$) and with very high IR luminosity surface brightness and gas temperatures (in excess of 200 K). Studies of IC~860 also find that the gas densities can be high with $n$$\sim$$10^7$ cm$^{-3}$ \citep{Aalto2019b}. 

Methyl cyanide CH$_{3}$CN, known as a tracer of dense and warm gas \citep[e.g.][]{Churchwell1983}, was detected in IC~860, NGC~4418, NGC~253 and NGC~4945. The line luminosity of CH$_{3}$CN 9-8 shows at least an order of magnitude higher values for NGC~4418 ($L_{\rm CH_{3}OH}=1.66\times 10^{5} L_{\odot}$) and IC~860 ($2.37\times 10^{5} L_{\odot}$) than for NGC~253 ($6.36\times 10^{3} L_{\odot}$) and NGC~4945 ($2.52\times 10^{4} L_{\odot}$). 

It seems likely that the CH$_{3}$CN emission is enhanced because of the high density and temperature in the CONs. The luminous CH$_{3}$CN emission in NGC~4418 may imply that the H$_{2}$S enhancement here is caused by desorption driven by radiation (Sec.~\ref{s:radiation}), for example through UV radiation from a buried source. However, NGC~4418 has an extremely large H$_2$ column density with $N$(H$_2$)$>10^{25}$ cm$^{-2}$ \citep{Costagliola2013,Sakamoto2013}. Since UV photons cannot penetrate long distances through such a high column density region, the global H$_{2}$S enhancement is unlikely to be caused by photodesorption by UV radiation. In addition, \citet{Bayet2009} argued that H$_{2}$S is unlikely to trace gas in Photon Dominated Regions (PDRs) in starburst galaxies, or near AGNs. 

Alternatively,  high energy photons and particles can pass through high column density cores and may photo- or thermally desorb H$_{2}$S out from grain surface over wider regions. Observations with the Chandra X-ray satellite by \citet{Maiolino2003} do not show a clear  X-ray signature in NGC~4418. The extreme column densities of NGC~4418 are difficult to penetrate completely for X-rays. However, the X-rays reach further than UV and it is possible that they may impact the nuclear chemistry and provide large bulk gas temperatures that can result in an H$_{2}$S enhancement . \citet{Gonzalez-Alfonso2013} proposed that X-ray ionisation due to an AGN is responsible for the molecular ion chemistry in the centre of NGC~4418.

Of the four LIRGs that have elevated H$_{2}$S/HCN intensity ratios, one does not show any evidence of either a molecular outflow or a "hot core": NGC~4826. This galaxy shows strong evidence of counter-rotating gas and stars, and its inner 1 kpc show the signatures of different types of $m$=1 perturbations (such as streaming motions) in the gas \citep{Garcia-Burillo2003}. A possibility is that the perturbations drive turbulence that may also lead to H$_{2}$S enhancements (Sec.~\ref{s:shocks}).

\subsection{Mass outflow rates and the global dense gas properties}
\label{s:outflow_correlations}

In Sec.~\ref{s:outflow1} we find no link between global H$_{2}$S abundance enhancements in the dense gas, and the presence (or velocity) of an outflow. There appears to be no general process that leads to significant H$_{2}$S abundance enhancements within the outflows, or when outflows are impacting their host galaxies.  (Although enhancements are still possible on smaller scales that are diluted in our single dish beam. )

However, to understand the launch mechanism and outflow evolution, it is also important to study and compare other molecular properties of the host galaxy to that of its outflow.  A correlation between the outflow mass, or its velocity, with for example the molecular gas mass or dense gas content of the host galaxy, gives essential clues to the origin of the molecular gas in the outflow. 

        To search for such host galaxy-outflow relations, we investigated correlations between the global line luminosity of H$_{2}$S, HCN, HCO$^+$ and the mass of the global molecular gas for our sample galaxies and the velocities and molecular masses of their outflows \citep[][Table 2]{Fluetsch2019}. The galaxies which are in our sample and also have their outflow information are Circinus, NGC~3256, NGC~1068, NGC~253, NGC~4418 and Mrk~231.

We find a correlation between $L$(H$_{2}$S) and the molecular mass, $M_{\rm outflow}$(H$_2$), of the outflow (Fig.\ref{fig:L_Mof}), where $L$(H$_{2}$S) is highest for the galaxies with the highest $M_{\rm outflow}$(H$_2$). This could be because $L$(H$_{2}$S) is a rough probe of the host galaxy molecular mass and that the two are connected,  i.e. that the ejected molecular gas mass is directly drawn from the large-scale molecular reservoir of the galaxy.  This notion can be tested through investigating the correlation between  $M_{\rm outflow}$ and other tracers of host galaxy molecular mass, such as the CO 1--0 luminosity, $L$(CO).

 We use global CO 1—0  luminosities, \citep{Sanders1991, Planesas1989, Aalto1995, Elmouttie1998, Houghton1997} for our sample galaxies, and find that $L$(CO) correlates more poorly with $M_{\rm outflow}$(H$_2$) than $L$(H$_{2}$S): the correlation coefficient is r=0.93 for H2S and r=0.57 for CO. : the correlation coefficient is $r$=0.93 for H$_{2}$S and $r$=0.50 for CO. The corresponding correlation coefficients for $L$(HCN) and $L$(HCO$^+$) are $r$=0.91 and $r$=0.83, respectively.  If we compare the critical densities, $n_{\rm crit}$, of the observed transitions, we find that H$_{2}$S and HCN have the highest $n_{\rm crit}$ followed by HCO$^+$ while the 1--0 transition of CO has an $n_{\rm crit}$ several orders of magnitude lower. {\it The molecular mass in the outflow appears to be more strongly connected to the dense gas (traced by H$_{2}$S and HCN) in the host galaxy, rather than the total available molecular reservoir.} Note that this suggested correlation between H$_{2}$S (and HCN) and the outflow mass, is different from the lack of correlation between H$_{2}$S abundances and outflow properties discussed in Sec~\ref{s:shocktest}. Here we use H$_{2}$S as a probe of the densest molecular gas, instead of searching for enhancements in the H$_{2}$S abundances as an effect of the outflow.

\begin{figure}
	\begin{center}
		\includegraphics[angle=0,width=0.5\textwidth]{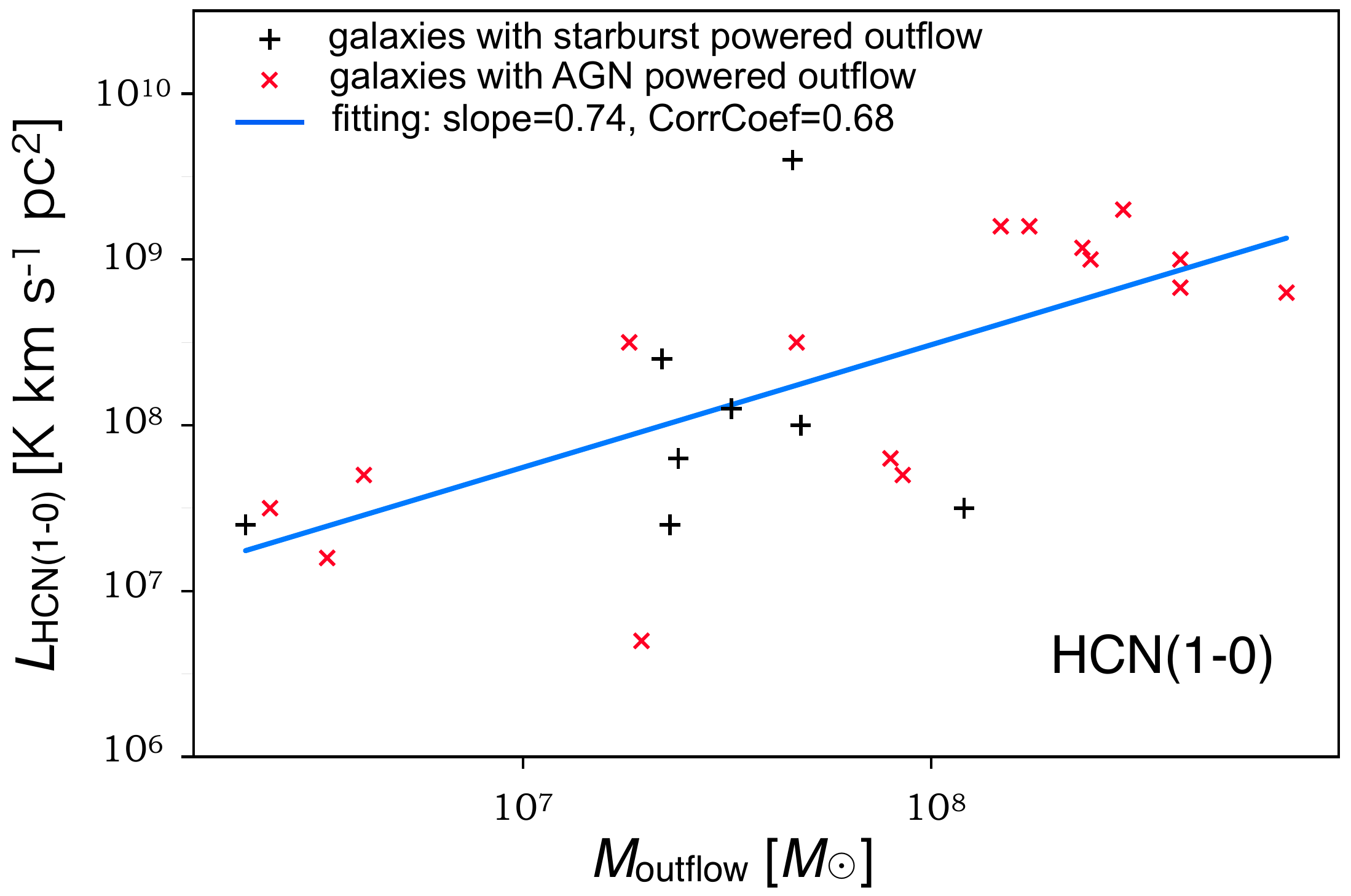}
		\caption{Same plot as Fig.\ref{fig:L_Mof} of the luminosities of HCN 1--0. The values for the molecular mass of the outflow, $M_{\rm outflow}$, are from \citet{Fluetsch2019}. $L_{\rm HCN 1-0}$ are from the references in Table \ref{tab:HCNlit}. AGN powered outflows are indicated with the red symbol x.} \label{fig:Mout_Lhcn}
	\end{center}
\end{figure}

Since the correlations are found for the relatively limited number of  sources in our sample, we decided to search for HCN  data in the literature to test the notion that $M_{\rm outflow}$(H$_2$) is more strongly correlated with the dense gas tracers than CO 1--0.  We found HCN 1–0 data for 24 galaxies in the \citet{Fluetsch2019} sample (listed in table \ref{tab:HCNlit}). HCN 1—0 has a lower critical density than the 2—1 transition and we expect to see a lower correlation coefficient with $M_{\rm outflow}$(H$_2$) than for HCN 2—1, which we also find with a value of $r$=0.68 (Fig.\ref{fig:Mout_Lhcn}). \citet{Fluetsch2019} also lists values for $M$(H$_2$) for their sample outflow host galaxies. Using these values for the 24 galaxies for which we have HCN 1—0 information, we find a correlation coefficient between $M$(H$_2$) and $M_{\rm outflow}$(H$_2$)  of $r$=0.25. This is significantly lower than what we find for $L$(CO) for our sample galaxies (Fig. 4). This may be because the molecular masses listed in \citep{Fluetsch2019} are not global, masses, or because the the correlation between $M_{\rm outflow}$(H$_2$) and the total reservoir becomes even poorer in large samples.

\begin{table}
	
	\caption{HCN 1-0 data from the literature\label{tab:HCNlit}}
	\begin{center}
		\begin{tabular}{lccc}
			\hline
			Source & Log($M_{\rm out}$) & Log($L_{\rm HCN})$ & Reference\\
			& [M$_{\odot}$] & [K km~s$^{-1}$ pc$^2$] &\\
			\hline
			\\
			IRAS F08572+3915 & 8.61 & 8.83 & (1) \\
			IRAS F10565+2448 & 8.37 & 9.07 & (2) \\
            IRAS 23365+3604 & 8.17 & 9.2 & (2) \\
            Mrk 273 & 8.24 & 9.2 & (2) \\
            I Zw 1 & 7.67 & 8.5 & (3) \\
            Mrk 231 & 8.47 & 9.3 & (4) \\
            NGC 1266 & 7.93 & 7.7 & (2) \\
            M82 & 8.08 & 7.5 & (5) \\
            NGC 1377 & 7.29 & 6.7 & (2) \\
            NGC 6240 & 8.61 & 9.0 & (4) \\
            NGC 3256 & 7.34 & 8.4 & (2) \\
            NGC 3628 & 7.36 & 7.4 & (2) \\
            NGC 253 & 6.32 & 7.4 & (2) \\
            NGC 6764 & 6.53 & 7.2 & (6) \\
            NGC 1068 & 7.26 & 8.5 & (2) \\
            NGC 2146 & 7.68 & 8.0 & (2) \\
            IRAS 17208-0014 & 7.66 & 9.6 & (4) \\
            NGC 1614 & 7.51 & 8.1 & (5) \\
            Circinus Galaxy & 6.38 & 7.5 & (7) \\
            NGC 1808 & 7.38 & 7.8 & (7) \\
            M51 & 6.61 & 7.7 & (5) \\
            PG 0157 + 001 & 8.39 & 9.0 & (3) \\
            IRAS 05189-2524 & 8.87 & 8.8 & (5) \\
            NGC 4418 & 7.90 & 7.8 & (8) \\
			\hline
		\end{tabular}
	\end{center}
	\tablefoot{References:The mass of the outflow are taken from Table 2 of \citet{Fluetsch2019}. The line luminosity:(1)\citet{Imanishi2007}, (2)\citet{Huang2018}, (3)\citet{Evans2006}, (4)\citet{Gracia-Carpio2006}, (5)\citet{Gao2004}, (6)\citet{Contini1997}, (7)\citet{Curran1999}, (8)\citet{Costagliola2011}}
\end{table}%

A possible interpretation is that the molecular mass in the outflow is linked to the properties and mass of the most centrally concentrated gas, traced by HCN and H$_{2}$S. This is expected if the outflows are launched from the inner regions of galaxies. In addition, it also compatible with the scenario where the gas is being ejected from the host galaxy in molecular form (through entrainment by jets and winds, a molecular MHC (Magneto Hydrodynamic) wind, or by radiation pressure \citep[e.g][]{Veilleux2013}). It appears less consistent with the notion that the molecular gas is primarily formed in the outflows themselves through instabilities in the hot gas  \citep[e.g.][]{Zubovas2014,Ferrara2016}. 

\citet{Lutz2020} find a correlation between bolometric luminosity, $L$(bol), and $M_{\rm outflow}$(H$_2$), with similar  relations that we find between $L$(HCN) and $M_{\rm outflow}$(H$_2$). They also find that the AGN luminosity, $L$(AGN), correlates  with $M_{\rm outflow}$(H$_2$) for objects with $L$(AGN)$>$$10^{10}$ L$_{\odot}$. \citet{Lutz2020} point out that $L$(bol) is linked to both AGN activity and star formation, making the relation to $L$(bol) somewhat difficult to interpret, but that AGN activity plays and important role in outflow driving. We note that the link between $L$(HCN) and $M_{\rm outflow}$(H$_2$) is stronger for the objects that \citet{Fluetsch2019} identify as AGN powered (see Fig.~\ref{fig:Mout_Lhcn}). 

The correlation between $M_{\rm outflow}$ and luminosity of dense gas tracers in the host galaxy should be tested on larger samples.  Furthermore, high resolution imaging is required to test the notion of a higher degree of central concentration for H$_{2}$S and HCN and to what degree this emission is linked to the launch regions of outflows.

\subsection{H$_{2}$S and H$_2$O}

\begin{figure*}
	\begin{center}
		\includegraphics[angle=0,scale=1]{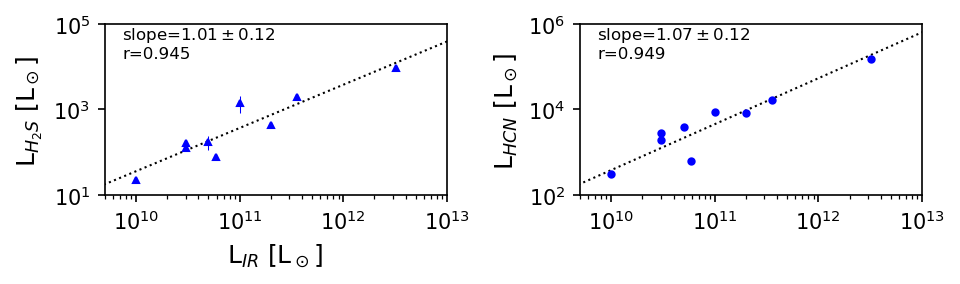}
		\caption{Line Luminosity of H$_{2}$S (left) and HCN (right) versus Infrared luminosity of each sample galaxy in which H$_{2}$S line is detected. 
		The correlation coefficient and the slope of the fitting line are given in the upper left corner of each panel.}\label{fig:LLumiH2S}
	\end{center}
\end{figure*}


Because of its chemical similarity, H$_{2}$S shares similar formation processes as H$_2$O, including the grain chemistry where both molecules may go through thermal- or photo- (non-thermal) desorption by shock or radiation. Although the more volatile H$_{2}$S will leave the grains at lower temperatures than more tightly bound H$_2$O \citep{Rodgers2003}. The discussion below is under two assumptions: 1) sulphur hydrogenates in H$_{2}$S on the dust grains, which is still debated, and 2) H$_{2}$S does not get dissociated when it is desorbed, even though \citet{Oba2019} showed it actually does. Due to the similarities it is conceivable that H$_{2}$S may serve as a water proxy, with its ground-state line accessible by ground-based telescopes for nearby galaxies. In this case,  we would expect a correlation between the H$_2$O and H$_{2}$S properties of our sample galaxies and open up the possibility to use H$_{2}$S as an initial substitute in searches for water in nearby galaxies.

\citet{Yang2013} fitted several transition of water emission luminosities against the corresponding $L_{\rm IR}$ \citep[see][Fig. 1 and Table 1]{Yang2013}. The fit can be described as,
\begin{equation}
	{log L_{\rm H_{2}O}=\alpha log L_{\rm IR}+ \beta.}
	\label{linefit}
\end{equation}

\noindent
They reported the results for eight transitions and the factor $\alpha$ varied between 0.78 and 1.21. For the ground state (para) H$_2$O line they find a $\alpha$=0.89$\pm$0.09 (using $\chi^2$ fitting). We carried out the same fitting for H$_{2}$S and HCN. The line luminosities were calculated using the formula $L_{\rm line} [L_{\odot}] = 1.04 \times 10^{-3} \times S_{\rm line} \Delta v  D^{2}_{\rm L} \times \nu_{\rm obs} $, where $S_{\rm line} \Delta v$ is the measured flux of the line [Jy km s$^{-1}$], D$_{\rm L}$ is the luminosity distance [Mpc] and$\nu_{\rm obs}$ is the frequency of the line [GHz]. $I$[Jy km s$^{-1}$] = $I$[K km s$^{-1}$] $\times$ 40.67 [Jy K$^{-1}$] is used to calculate the line flux from the observed antenna temperature \citep{Carilli2013}\footnote{http://www.apex-telescope.org/telescope/efficiency/index.php?yearBy=2017}. In Fig. \ref{fig:LLumiH2S} we show plots of the line luminosity of H$_{2}$S and HCN versus the infrared luminosity of each galaxy. 

We found a factor $\alpha$ (in the Eq.(\ref{linefit})) of 1.01$\pm$0.12 ($\chi^2$) for H$_{2}$S and 1.07$\pm$0.12 ($\chi^2$) for HCN. This is within the error of the value found for the ground-state H$_2$O line. \citet{Yang2013} suggested that the reason that the correlation between H$_2$O and IR is near-linear is IR pumping of the H$_2$O lines. They argue that, after absorbing a far-IR photon, the upper levels cascade down to populate lower levels in a constant fraction, yielding a linear relationship with $L_{\rm IR}$, and that the linear correlation shows the importance of IR-pumping. Although it is true that IR-pumping is important for the excitation of H$_2$O it is not clear that the same logic can be applied to the near-linear correlations found for H$_{2}$S and HCN.  \citet{Crockett2014} claim that IR pumping of  H$_{2}$S for transitions lower than $J$=3 is not possible. The near-linear correlation between $L_{\rm HCN}$ and $L_{\rm IR}$ (first reported for LIRGs by \citep{Gao2004}) is also not clearly connected to IR pumping (even if the HCN rotational ladder can be pumped via a mid-IR bending mode \citep[e.g.][]{Aalto1994}). \citet{Gao2004} argue that the HCN-IR correlation is due to HCN probing dense gas engaged in star formation, that in turn gives rise to the IR luminosity. We conclude that H$_{2}$S, H$_2$O and HCN seem to correlate in a similar way with $L_{\rm IR}$ which may indicate that they emerge from similar regions in the LIRGs.



\section{Conclusions}
\label{s:conclusions}

Using the APEX telescope, we observed the $\lambda$=2~mm $1_{10}$--$1_{01}$ line of ortho-H$_{2}$S towards the centres of 12 nearby luminous galaxies and detected H$_{2}$S emission in nine of them. We also detected HCN and HCO$^+$ 2--1 in 11 of the sample galaxies and also HNC, SO, CH$_3$OH, CH$_3$CN, H$_2$CS and HOC$^+$ in some galaxies. In addition, we observed H$_{2}$S in the ULIRG Mrk~231 with the NOEMA telescope. Our aim was to study the impact of outflows on the chemistry and physical conditions of the gas in the host galaxy and outflow, using H$_{2}$S as a potential tracer of the impact of shocks. The line intensity ratio of H$_{2}$S to HCN (and to HCO$^+$) shows considerable variation, which could be due to significant changes of H$_{2}$S abundance among the studied galaxies. Four galaxies stand out to be particularly H$_{2}$S luminous: two of them exhibiting strong outflows, one harbouring a Compact Obsecured Nucleus (CON) and one shows evidence of gas and star counter-rotation possibly feeding turbulence. These results support previous suggestions that the H$_{2}$S emission may become enhanced in both shocked and irradiated dusty gas in the centres of luminous, gas-rich galaxies. However, for the sample galaxies as a whole, we find no correlation between the H$_{2}$S/HCN or the H$_{2}$S/HCO$^+$ line ratios with the presence of a molecular outflow or its speed. In Mrk~231 we detect H$_{2}$S out to -300 and +500 kms$^{-1}$, but the emission is not enhanced with respect to that of HCN. The relatively faint H$_{2}$S emission in the line wings appears inconsistent with the notion of shock-chemistry in the molecular outflow of Mrk~231.

We also investigated other possible connections between host galaxy molecular properties and that of their outflows. We found no link between the luminosities of H$_{2}$S $1_{10}$--$1_{01}$ , HCN 2--1 , HCO$^+$2--1   or CO 1--0  with the outflow velocity. In contrast, we did find a correlation with the molecular mass of the outflow, $M_{\rm outflow}$(H$_2$),  where the correlation coefficient is strongest for $L_{\rm H_{2}S}$ and $L_{\rm HCN}$ and weakest for $L_{\rm CO}$. We suggest that $L_{\rm H_{2}S}$ serves as a tracer of the dense gas content, similar to $L_{\rm HCN}$, and that the correlation between $L_{\rm H_{2}S}$ and $M_{\rm outflow}$(H$_2$) implies a relation between the dense gas reservoir and the properties and evolution of the molecular feedback. The dense gas component is likely more centrally concentrated than the bulk of the molecular gas (traced by CO 1--0) and a possible explanation is that the outflows are launched in the central regions and that the molecular gas in the outflow stems from the reservoir in the inner region, rather than being formed from instabilities in the hot gas. However, further studies of this correlation is important to understand its cause. In addition, it is important to mention that the enhancement of H$_{2}$S could also be related to starburst regions dominated by dense molecular gas and UV radiation.

Finally, based on chemical similarities between H$_{2}$S and H$_2$O, we studied if the two species show analogous behaviour in how they correlate to the IR luminosity.  We find that their IR-correlation coefficients are similar that could indicate that they originate in the same regions in the galaxy: warm gas in shocks or irradiated by star formation or an AGN.


\begin{acknowledgements}
      This research has made use of the services of the ESO Science Archive Facility. 
      
      Based on observations collected at the European Organisation for Astronomical Research in the Southern Hemisphere under ESO programmes 096.F-9331(A), 099.F-9312(A) and 0100.F-9311(A).
      
      Based on observations with the APEX telescope. APEX is a collaboration between the Max-Planck-Institut f{\"u}r Radioastronomie, the European Southern Observatory, and the Onsala Observatory.
      
      Based on observations carried out under project number W16BW with the IRAM NOEMA Interferometer. IRAM is supported by INSU/CNRS (France), MPG (Germany) and IGN (Spain).
      
      M.S., S.A. and S.K. gratefully acknowledges funding from the European Research Council (ERC) under the European Union’s Horizon 2020 research and innovation programme (grant agreement No 789410)
      
      SV acknowledges the European Research Council (ERC) Advanced Grant MOPPEX 833460. 
      
      YN is supported by NAOJ ALMA Scientific Research grant No. 2017-06B 
      and JSPS KAKENHI grant No. JP18K13577.
\end{acknowledgements}

\bibliography{Library_Mendeley.bib}

\begin{appendix}

\section{Spectra of the APEX sample galaxies.}\label{append1}
\begin{figure*}
		\begin{center}
			\begin{subfigure}{0.48\textwidth}
				\includegraphics[width=\linewidth]{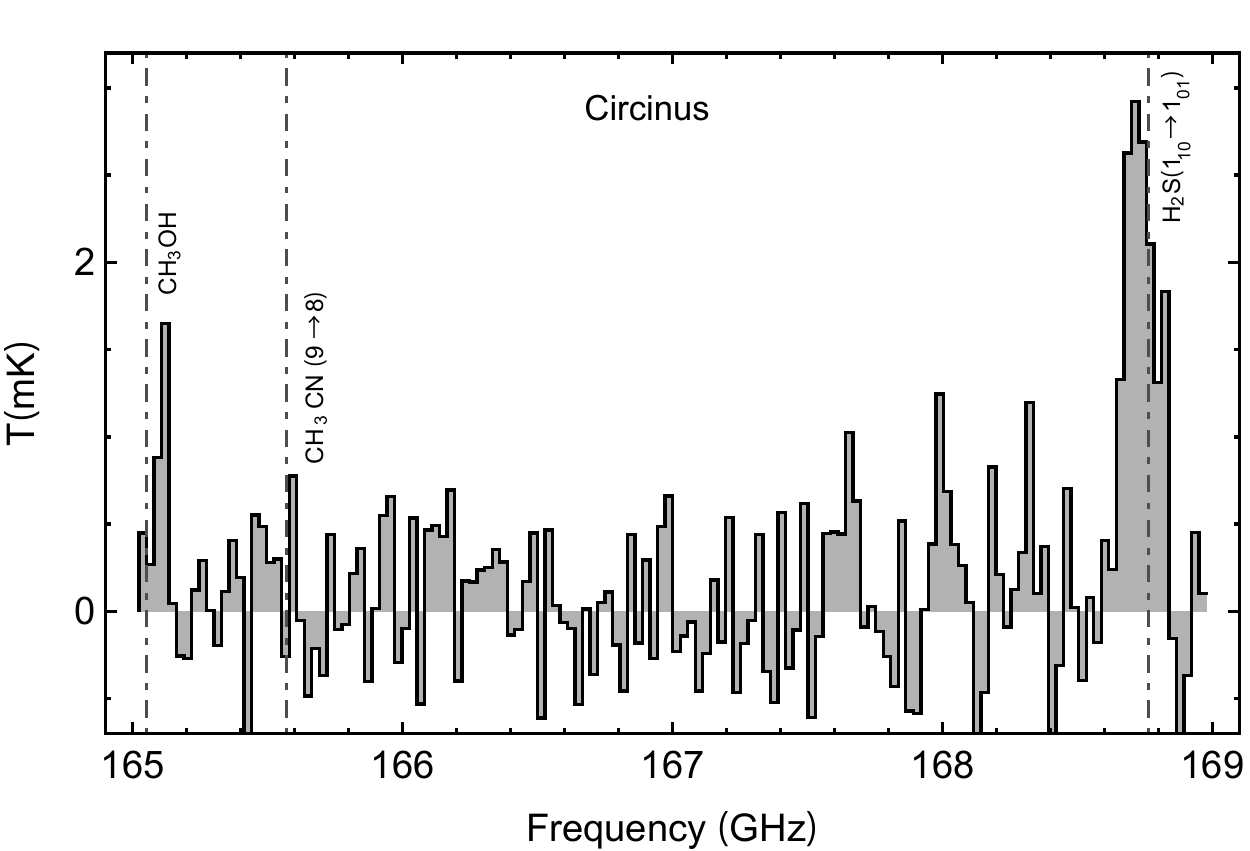}
				\caption{}\label{fig:a}
			\end{subfigure}
			\begin{subfigure}{0.48\textwidth}
				\includegraphics[width=\linewidth]{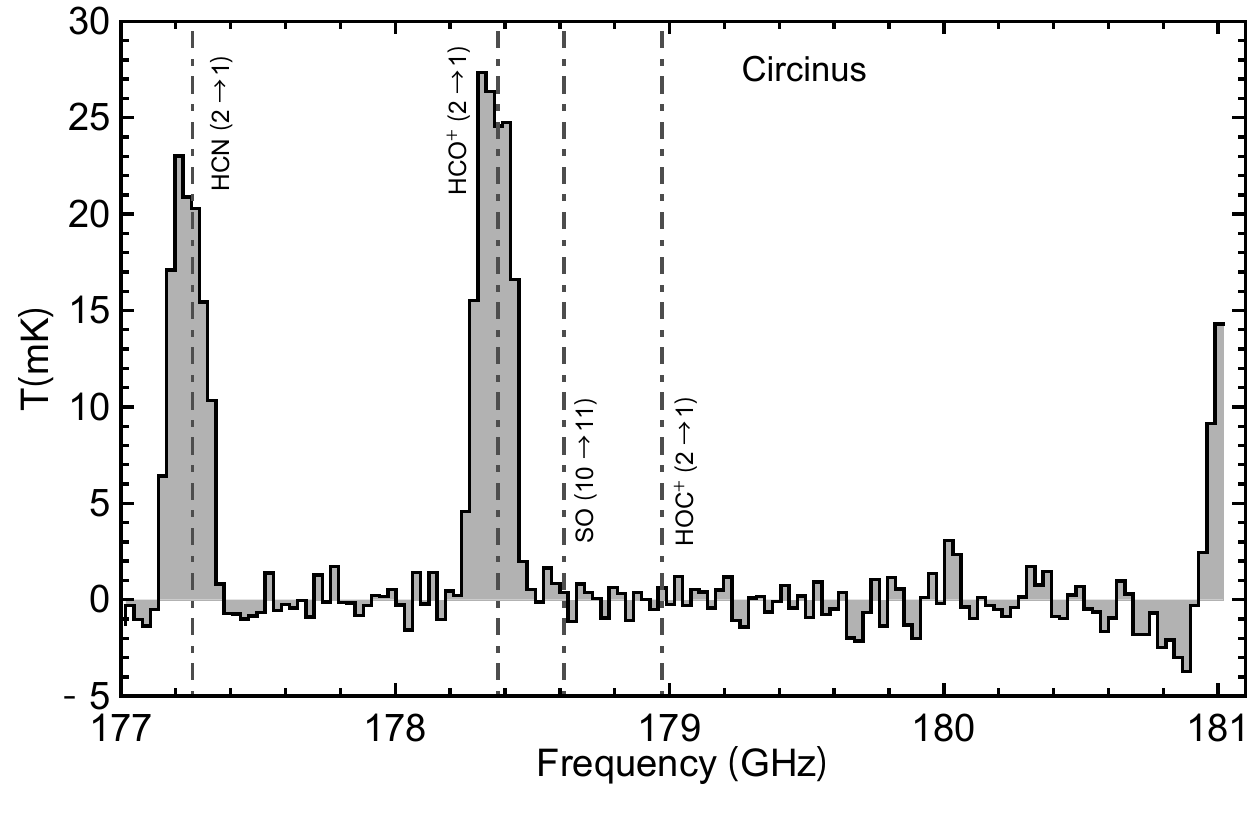}
				\caption{} \label{fig:b}
			\end{subfigure}
			\begin{subfigure}{0.48\textwidth}
				\includegraphics[width=\linewidth]{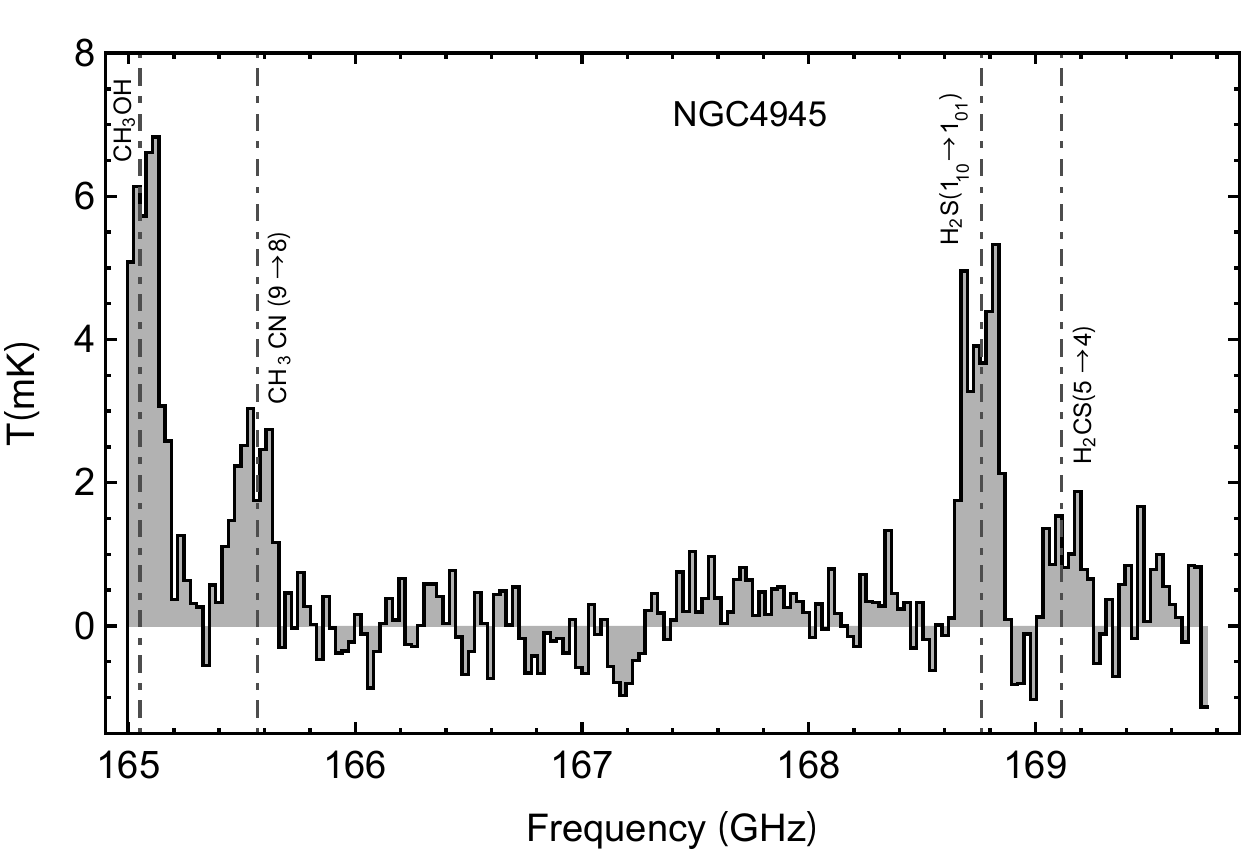}
				\caption{} \label{fig:c}
			\end{subfigure}
			\begin{subfigure}{0.48\textwidth}
				\includegraphics[width=\linewidth]{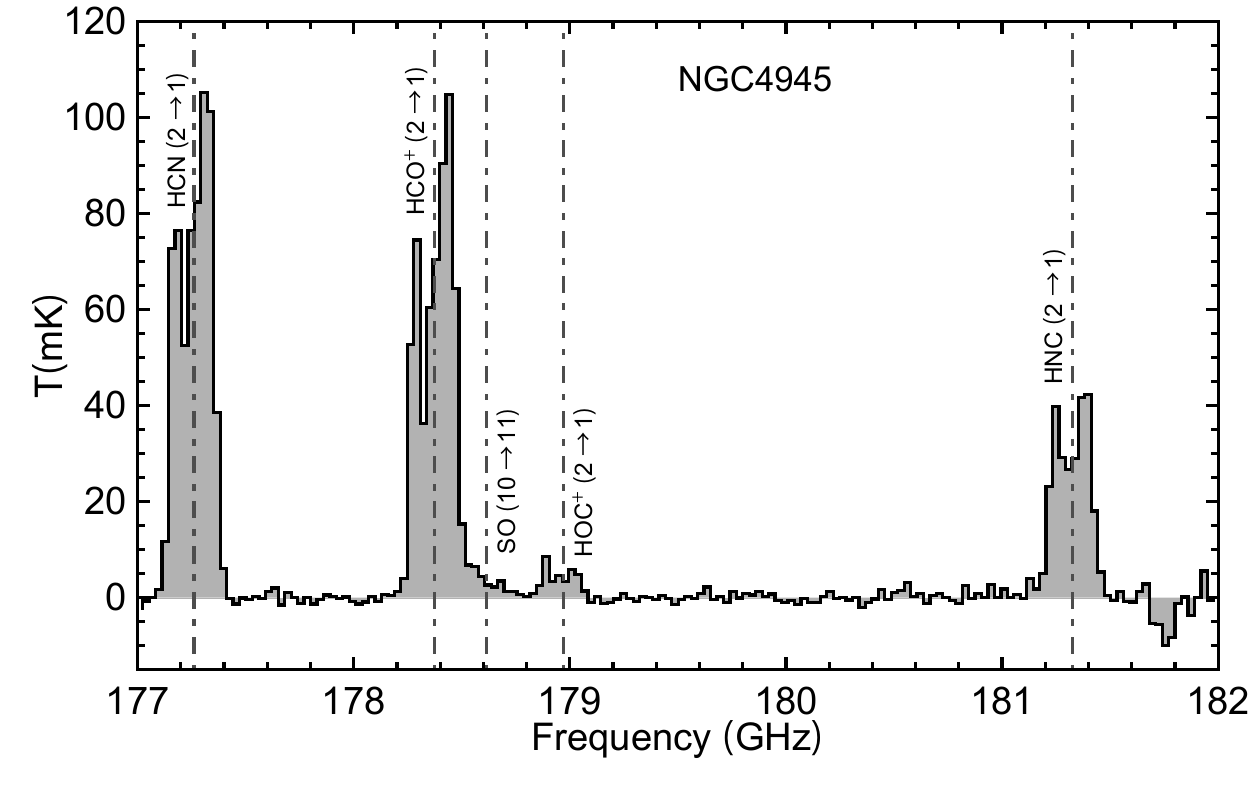}
				\caption{} \label{fig:d}
			\end{subfigure}
			\begin{subfigure}{0.48\textwidth}
				\includegraphics[width=\linewidth]{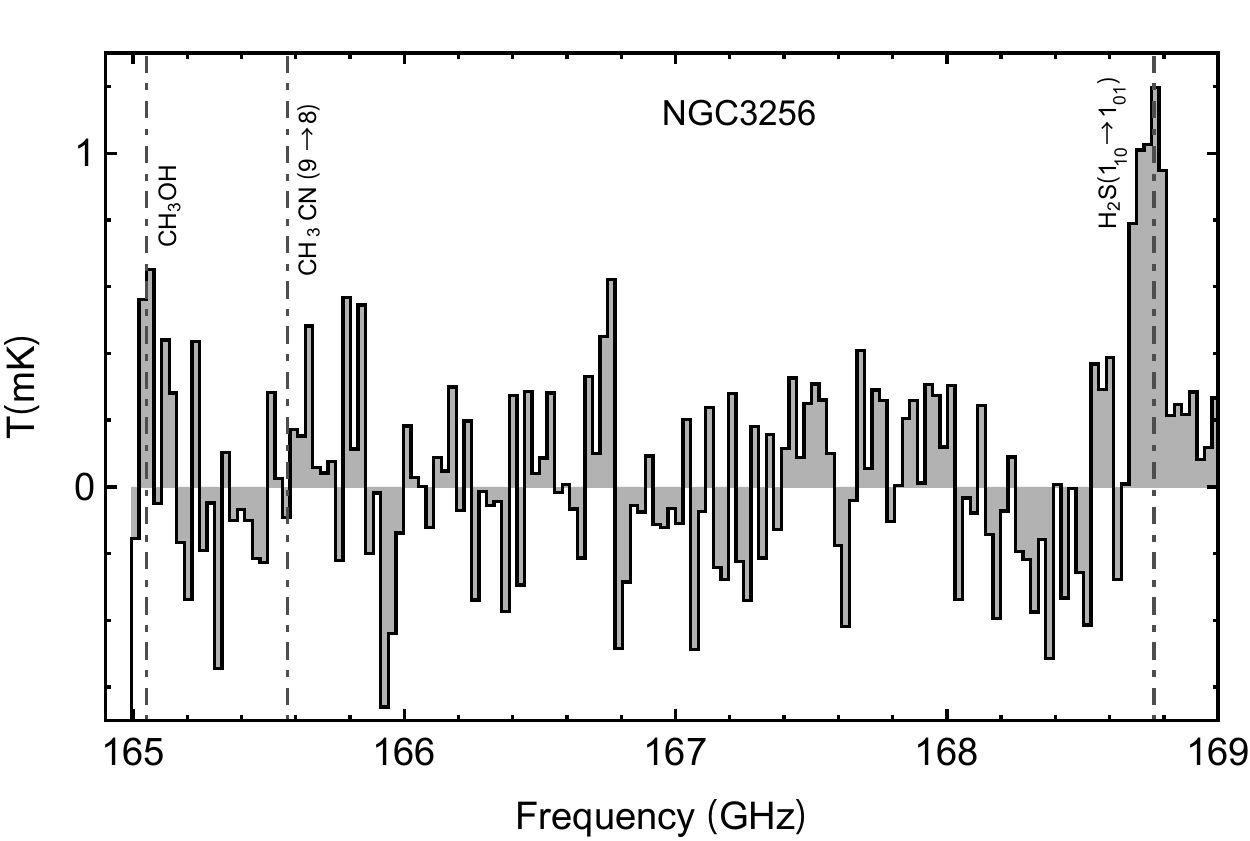}
				\caption{} \label{fig:e}
			\end{subfigure}
			\begin{subfigure}{0.48\textwidth}
				\includegraphics[width=\linewidth]{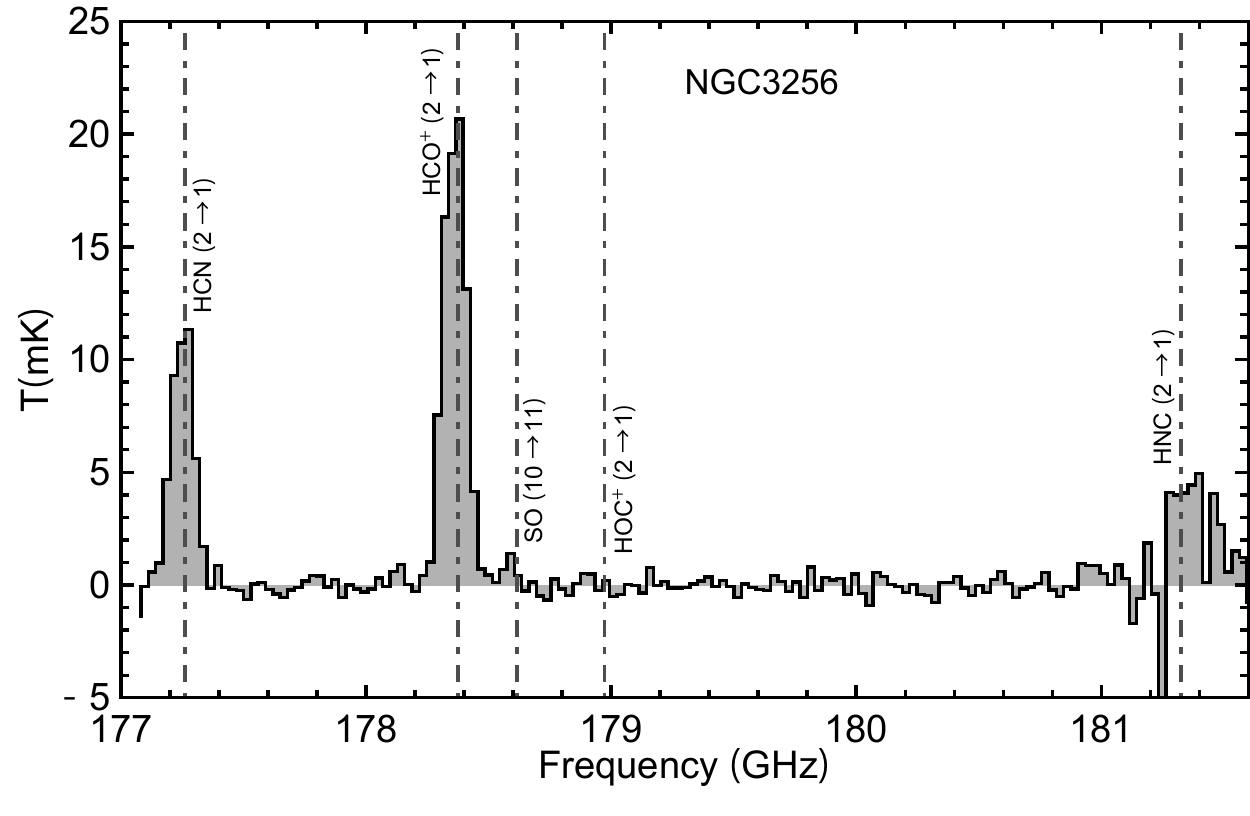}
				\caption{} \label{fig:f}
			\end{subfigure}
			\begin{subfigure}{0.48\textwidth}
				\includegraphics[width=\linewidth]{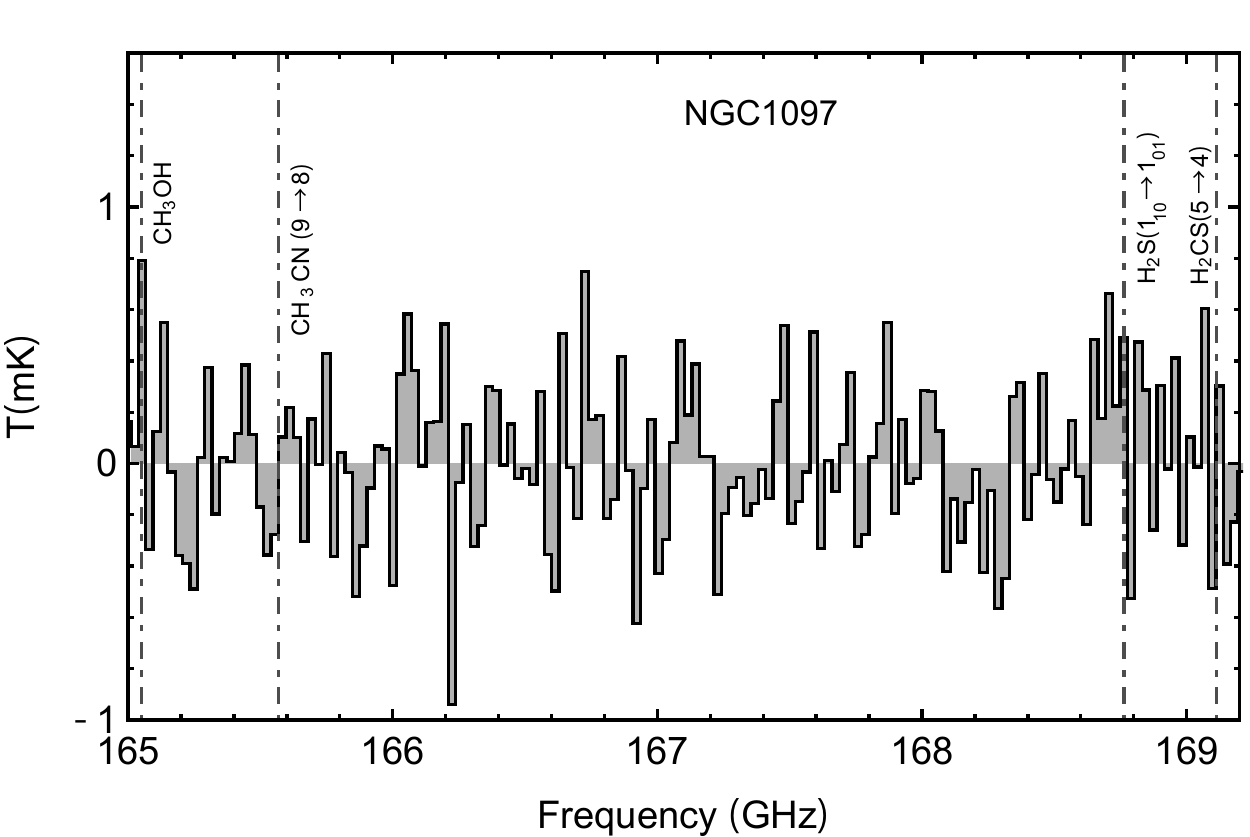}
				\caption{} \label{fig:g}
			\end{subfigure}
			\begin{subfigure}{0.48\textwidth}
				\includegraphics[width=\linewidth]{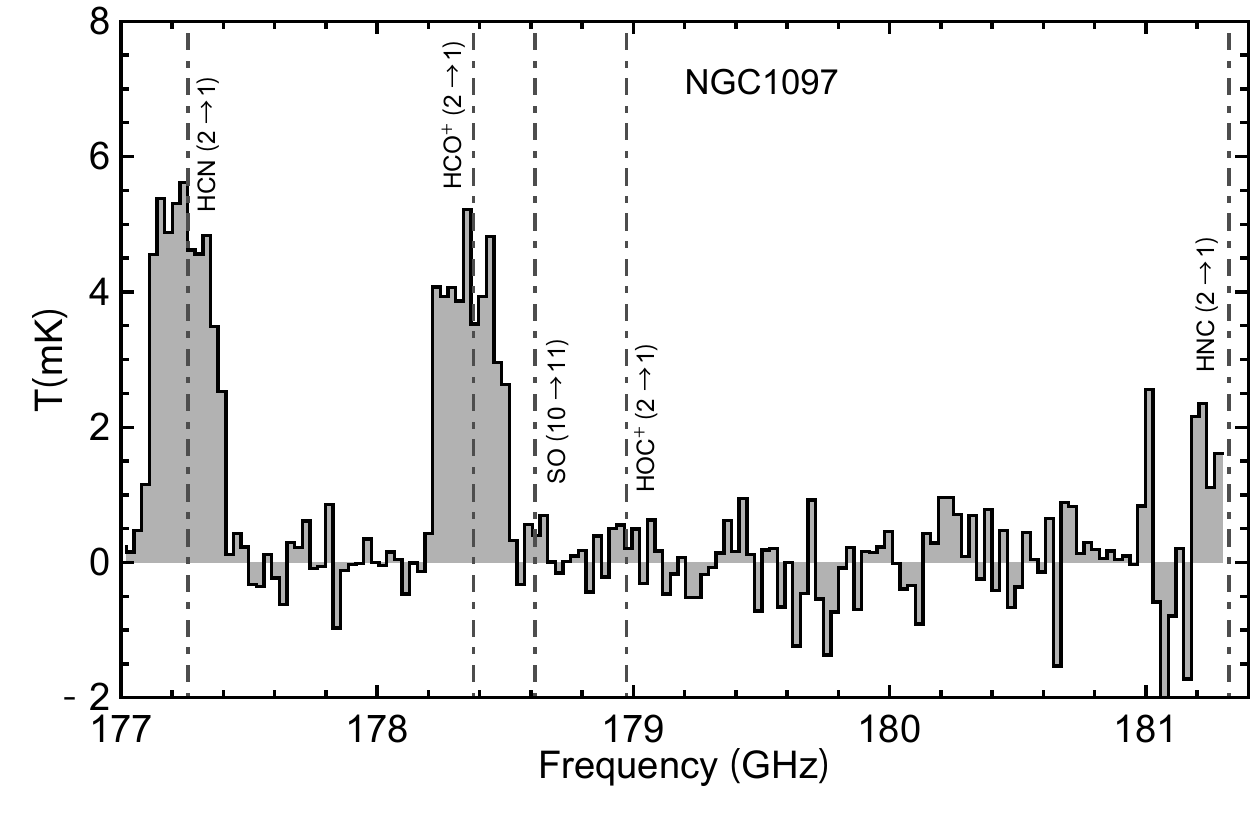}
				\caption{} \label{fig:h}
			\end{subfigure}
			\caption{\label{fig:spec1} Spectra taken with SEPIA. The intensity scale is in T$^{*}_{A}$, not corrected for the beam efficiency, which at this frequencies is 0.78. 
			}
		\end{center}
\end{figure*}
	
\begin{figure*}
		\ContinuedFloat
		\begin{center}
			\begin{subfigure}{0.48\textwidth}
				\includegraphics[width=\linewidth]{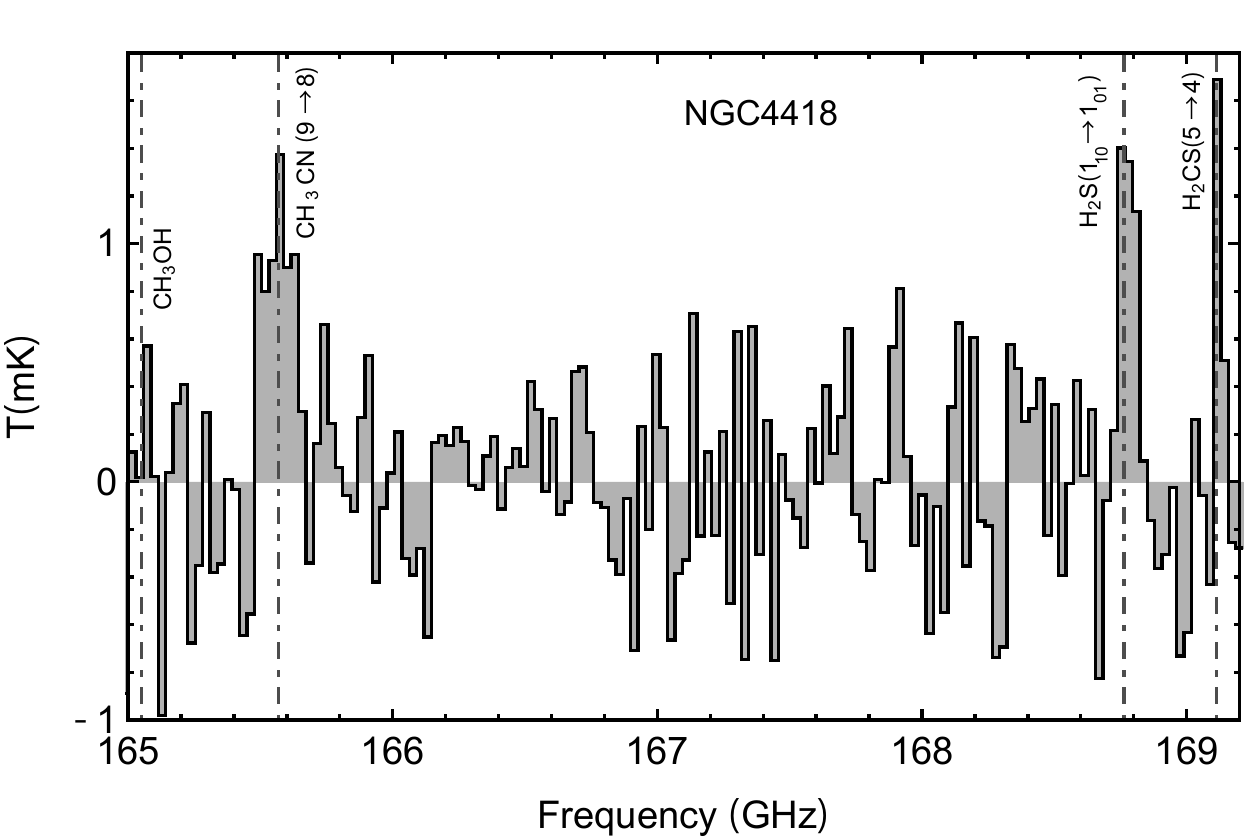}
				\caption{} \label{fig:i}
			\end{subfigure}
			\begin{subfigure}{0.48\textwidth}
				\includegraphics[width=\linewidth]{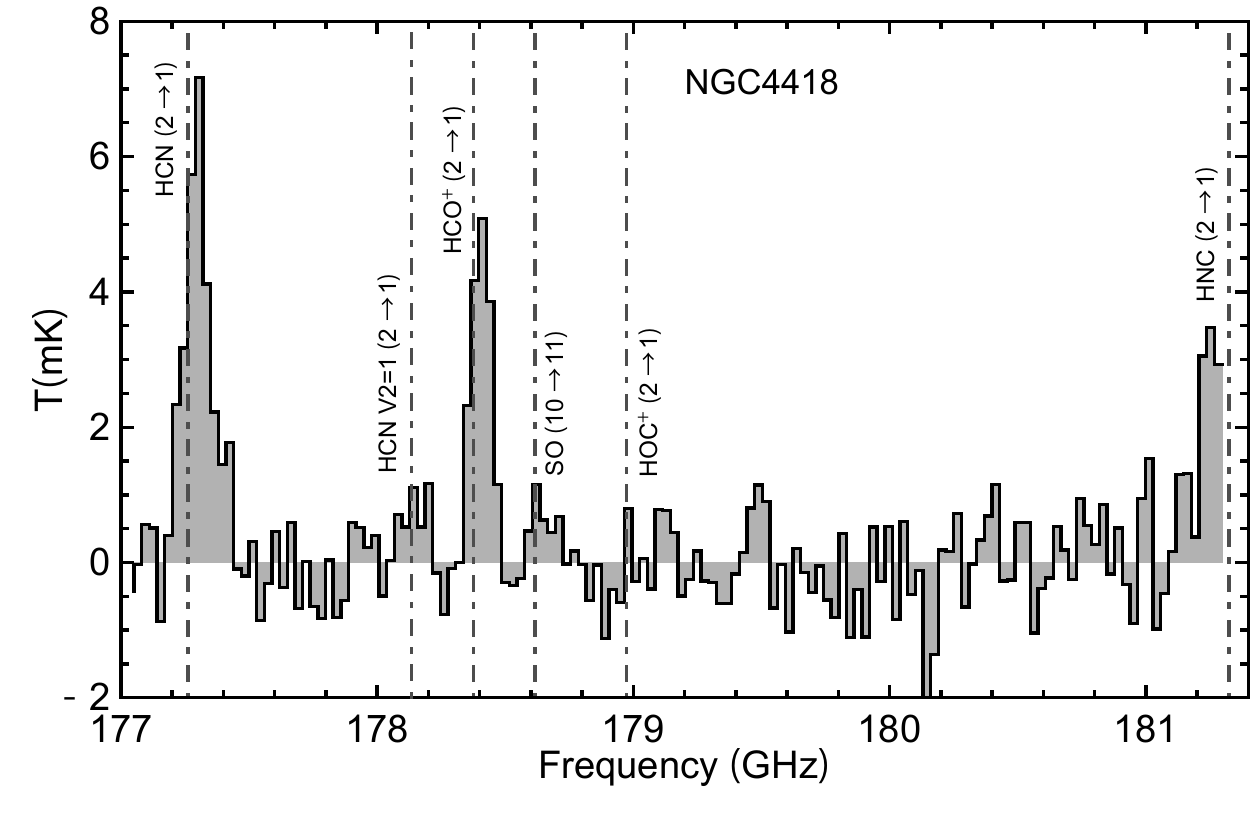}
				\caption{} \label{fig:j}
			\end{subfigure}
			\begin{subfigure}{0.48\textwidth}
				\includegraphics[width=\linewidth]{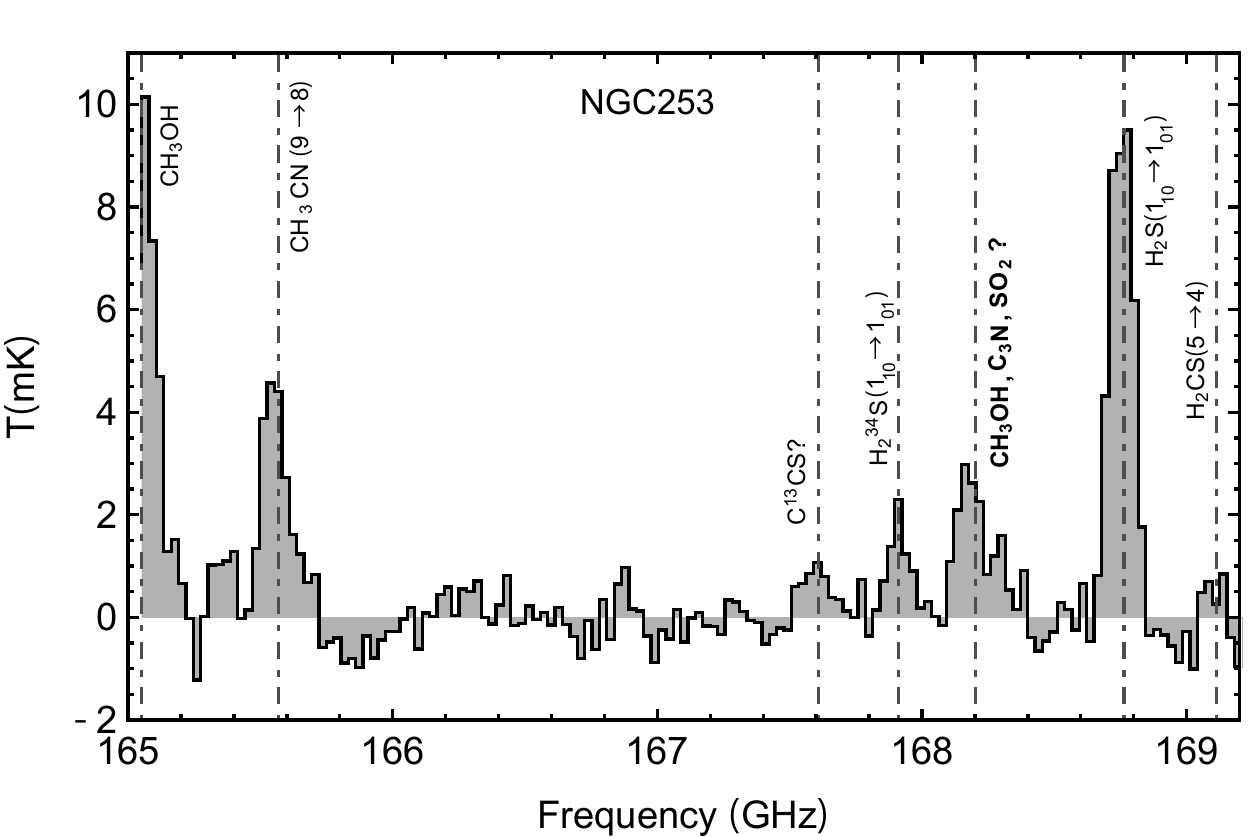}
				\caption{} \label{fig:k}
			\end{subfigure}
			\begin{subfigure}{0.48\textwidth}
				\includegraphics[width=\linewidth]{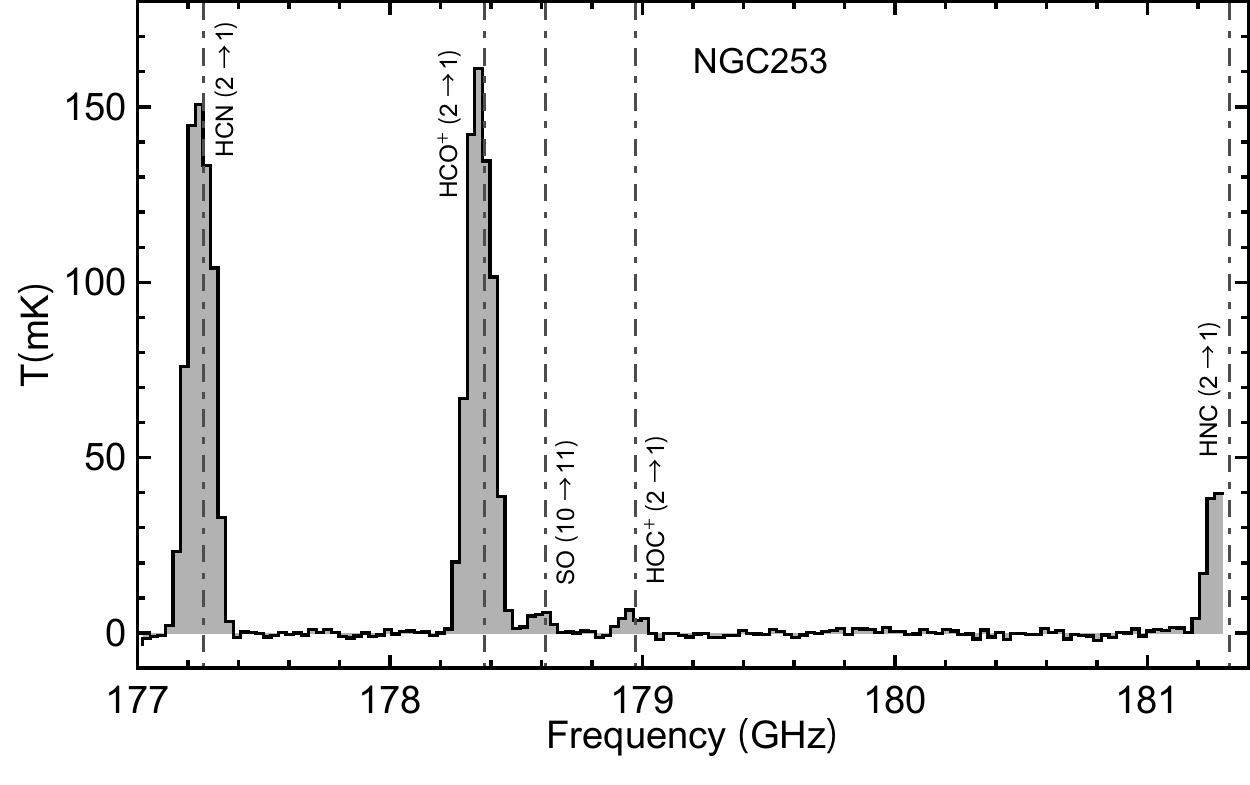}
				\caption{} \label{fig:l}
			\end{subfigure}	
			\begin{subfigure}{0.48\textwidth}
				\includegraphics[width=\linewidth]{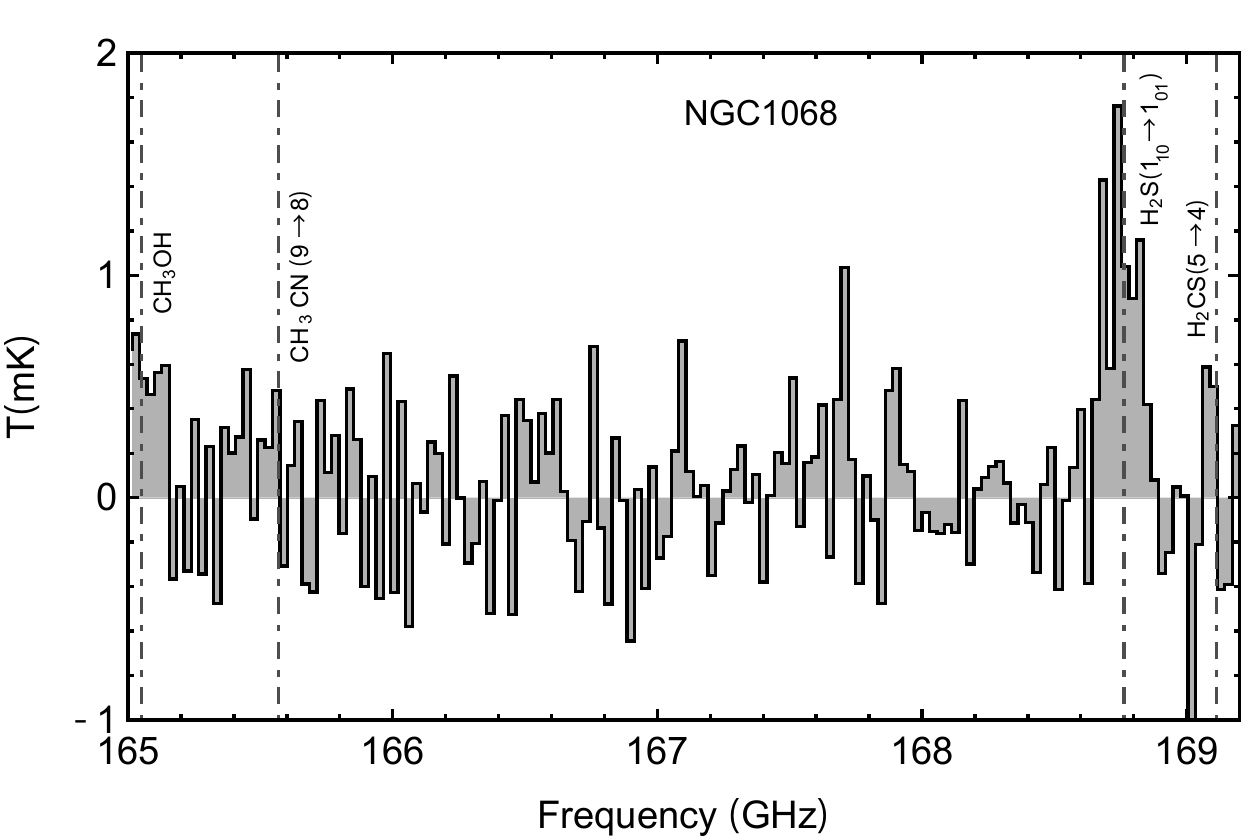}
				\caption{} \label{fig:m}
			\end{subfigure}
			\begin{subfigure}{0.48\textwidth}
				\includegraphics[width=\linewidth]{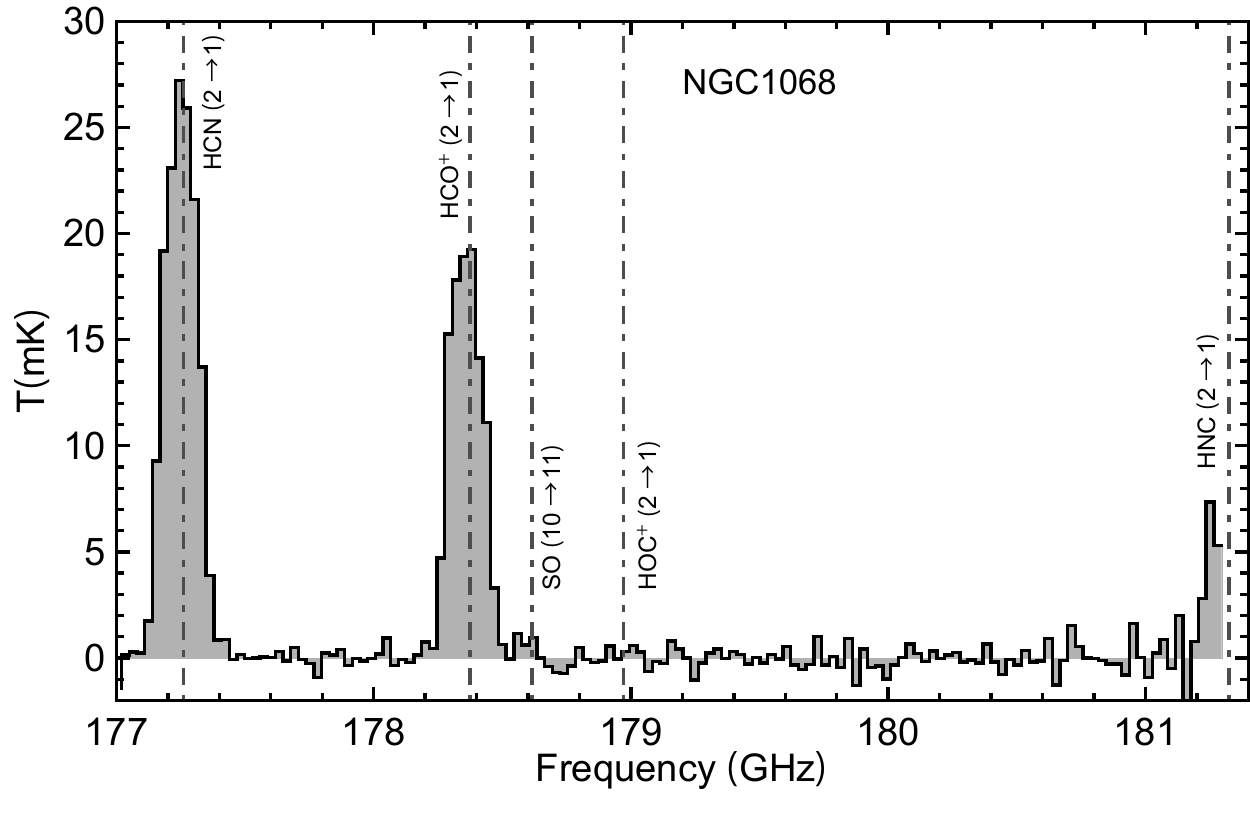}
				\caption{} \label{fig:n}
			\end{subfigure}
			\begin{subfigure}{0.48\textwidth}
				\includegraphics[width=\linewidth]{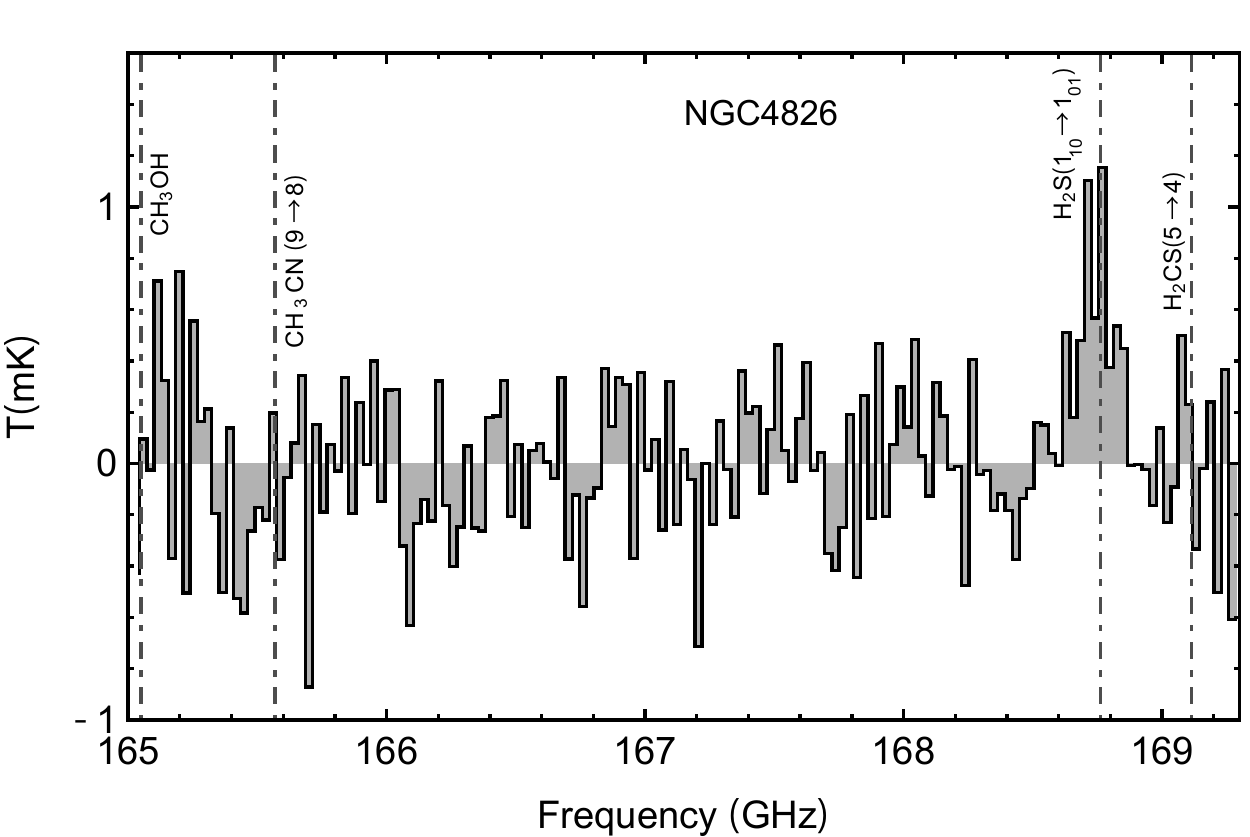}
				\caption{} \label{fig:o}
			\end{subfigure}
			\begin{subfigure}{0.48\textwidth}
				\includegraphics[width=\linewidth]{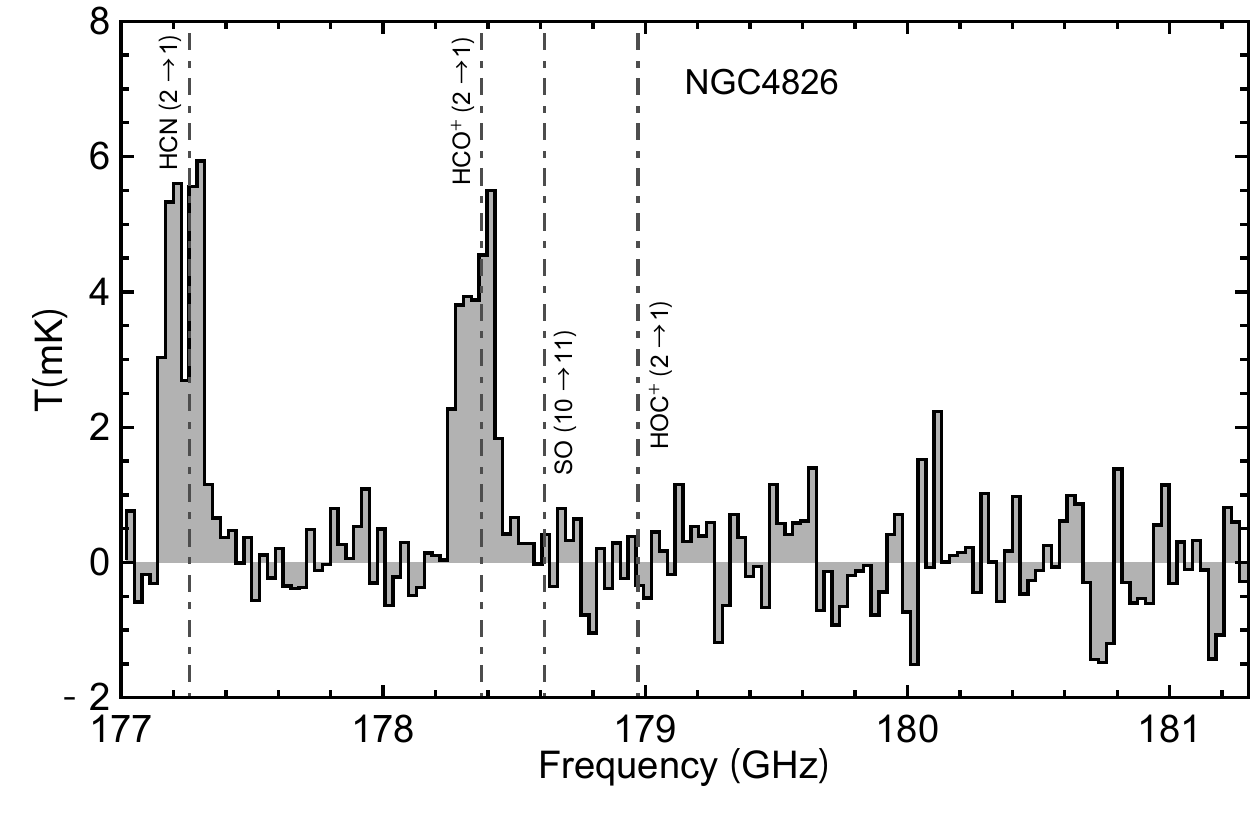}
				\caption{} \label{fig:p}
			\end{subfigure}
			\caption{\label{fig:spec2} (continued) 
			}
		\end{center}
\end{figure*}
	
\begin{figure*}
		\ContinuedFloat
		\begin{center}
			\begin{subfigure}{0.44\textwidth}
				\includegraphics[width=\linewidth]{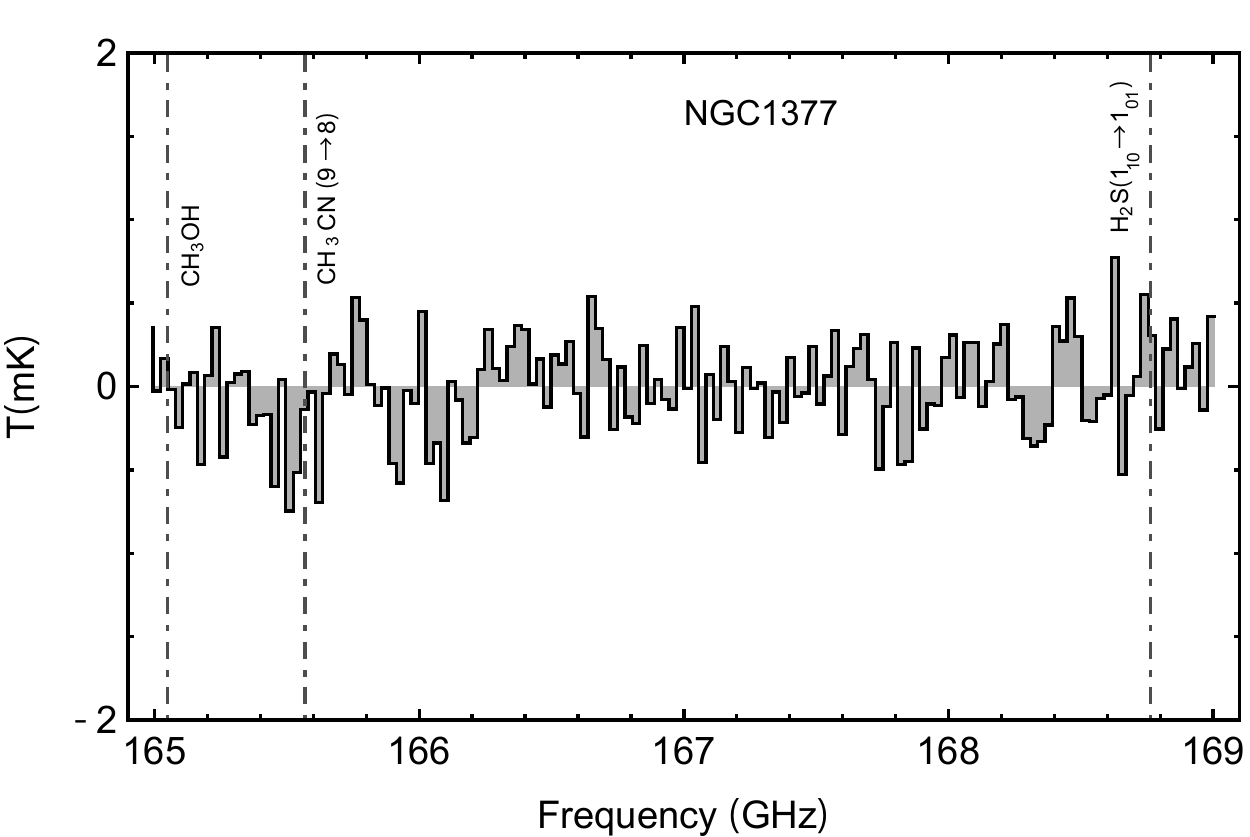}
				\caption{} \label{fig:q}
			\end{subfigure}
			\begin{subfigure}{0.44\textwidth}
				\includegraphics[width=\linewidth]{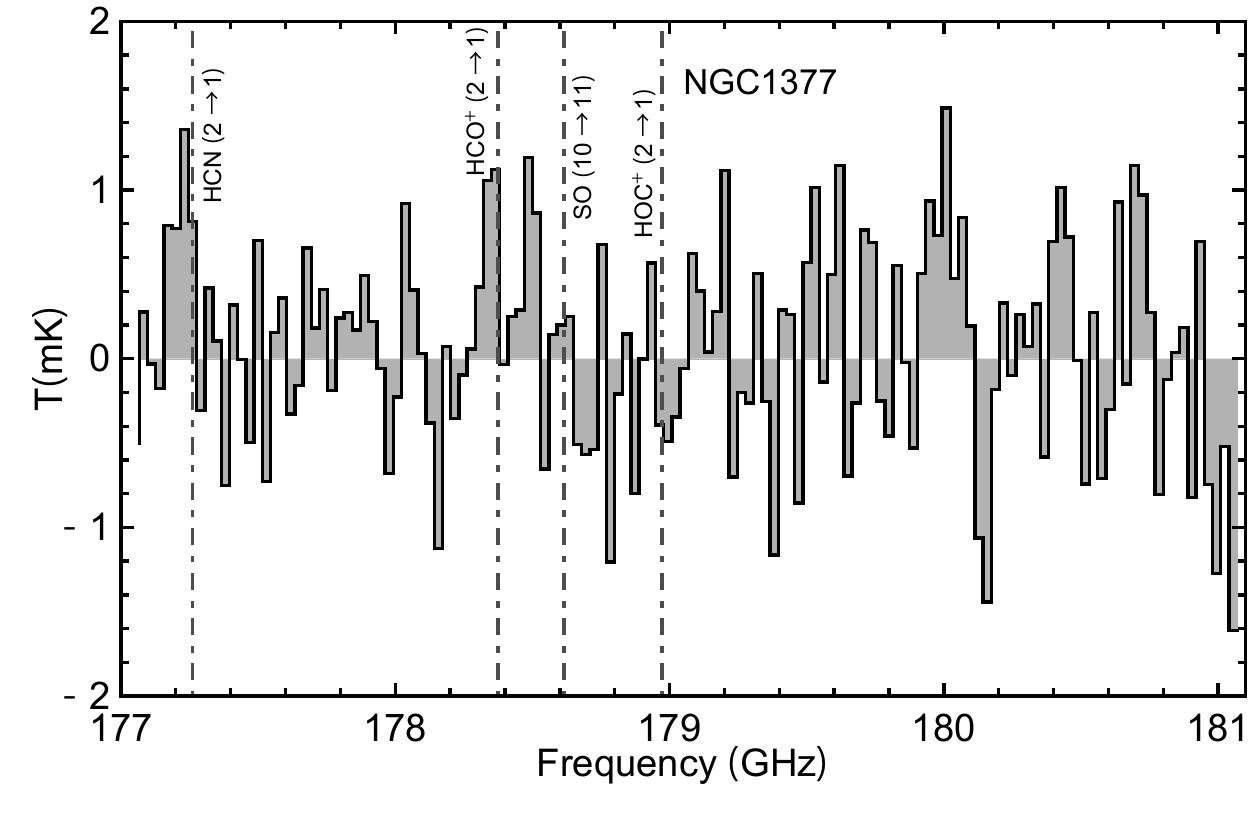}
				\caption{} \label{fig:r}
			\end{subfigure}
			\begin{subfigure}{0.44\textwidth}
				\includegraphics[width=\linewidth]{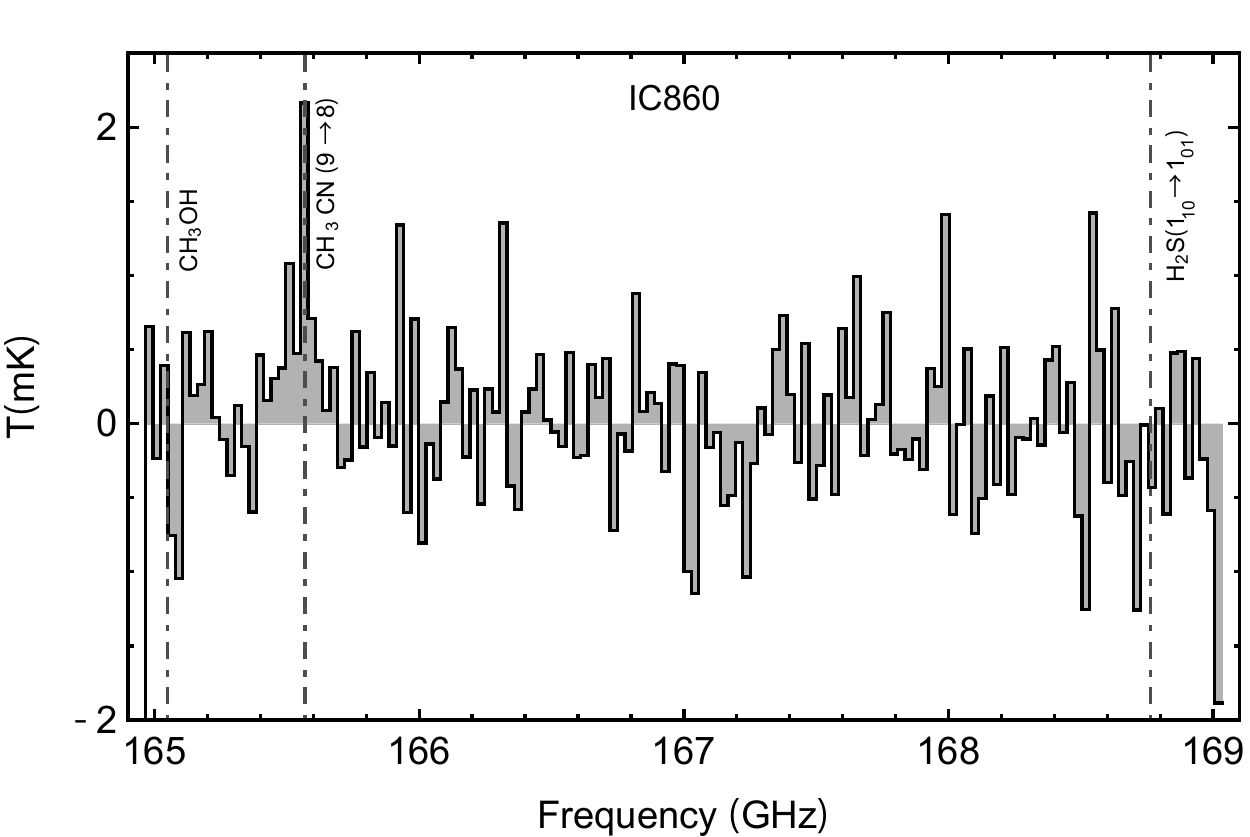}
				\caption{} \label{fig:s}
			\end{subfigure}
			\begin{subfigure}{0.44\textwidth}
				\includegraphics[width=\linewidth]{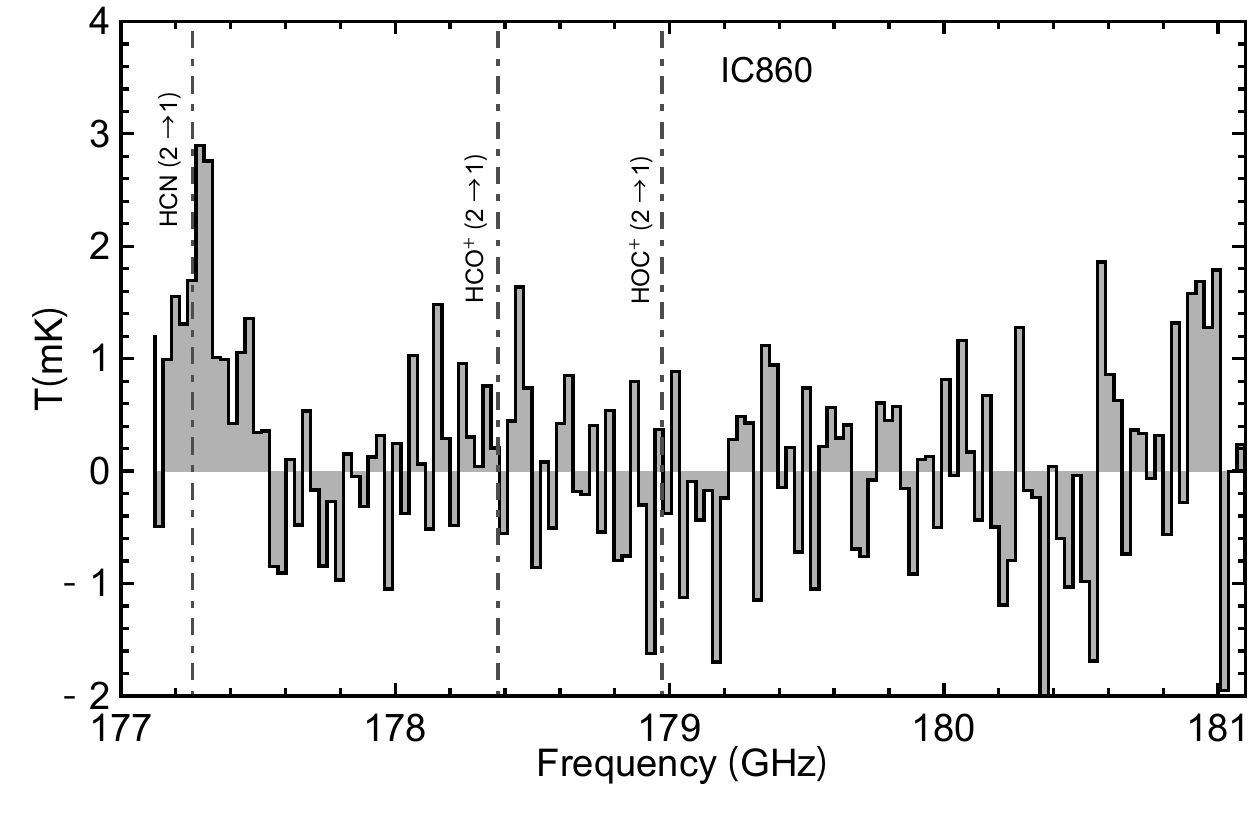}
				\caption{} \label{fig:t}
			\end{subfigure}
			\begin{subfigure}{0.44\textwidth}
				\includegraphics[width=\linewidth]{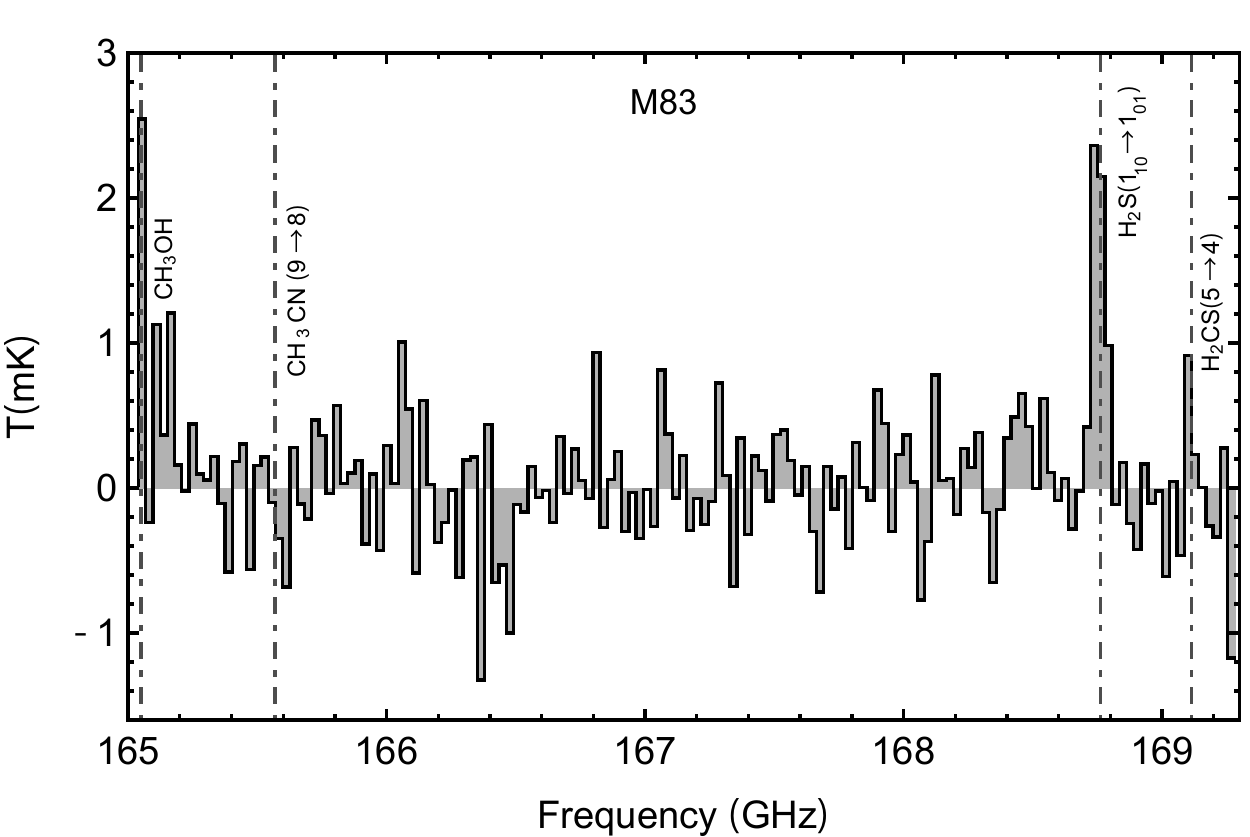}
				\caption{} \label{fig:u}
			\end{subfigure}
			\begin{subfigure}{0.44\textwidth}
				\includegraphics[width=\linewidth]{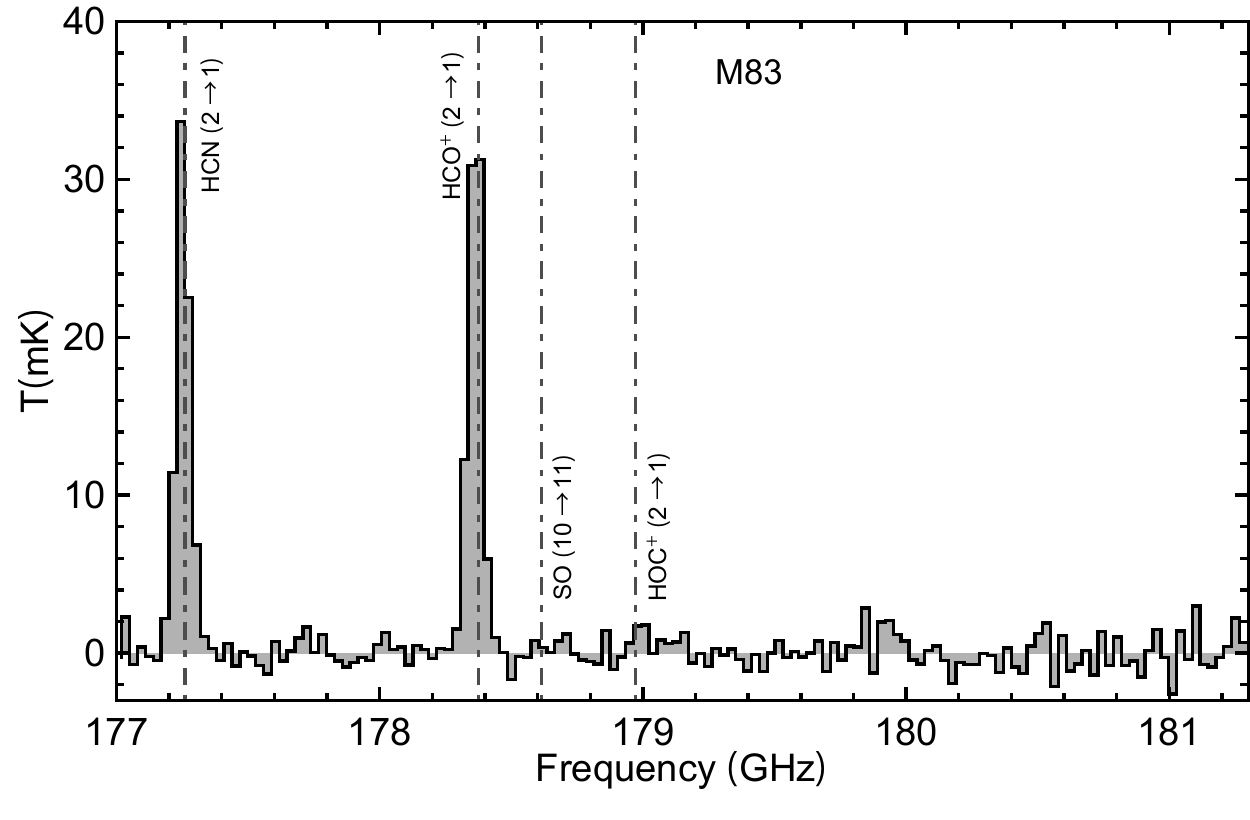}
				\caption{} \label{fig:v}
			\end{subfigure}
			\begin{subfigure}{0.44\textwidth}
				\includegraphics[width=\linewidth]{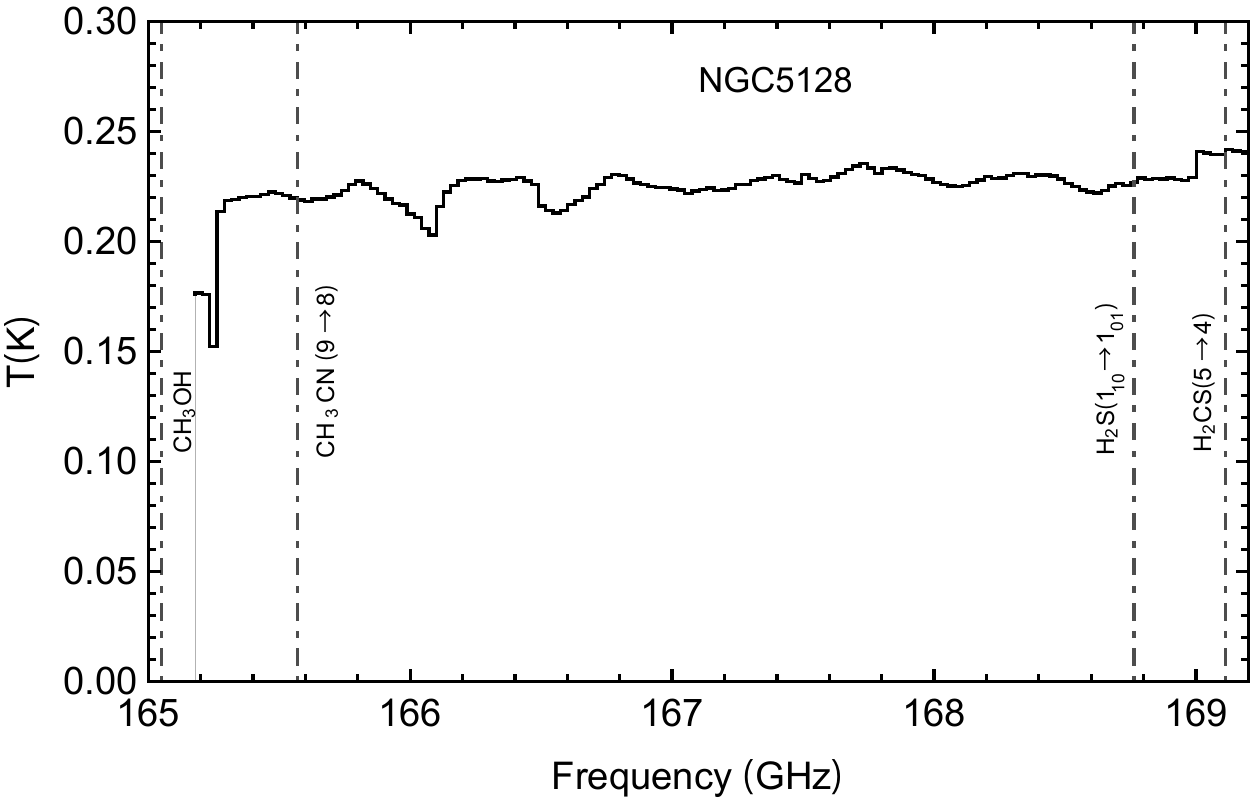}
				\caption{} \label{fig:w}
			\end{subfigure}
			\begin{subfigure}{0.44\textwidth}
				\includegraphics[width=\linewidth]{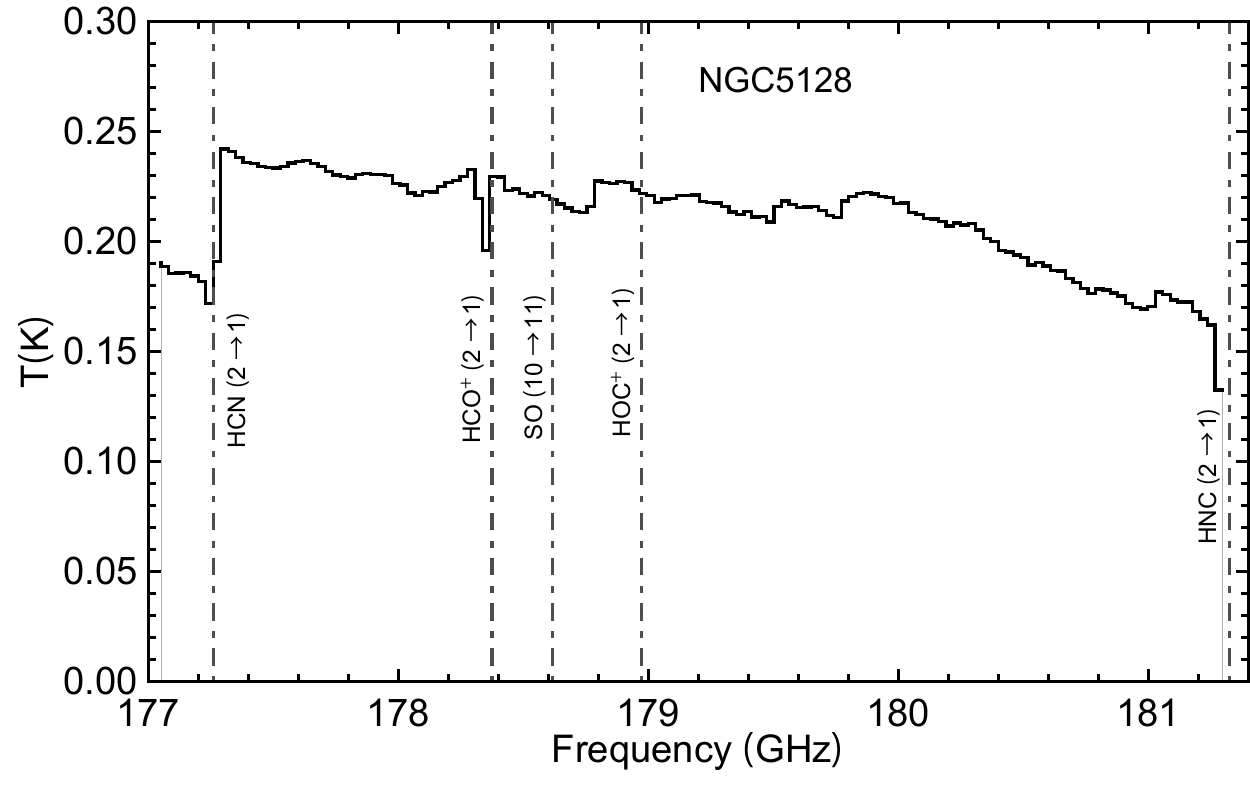}
				\caption{} \label{fig:x}
			\end{subfigure}
			\caption{\label{fig:spec3} (continued)
			}
		\end{center}
\end{figure*}

\section{APEX results}\label{tab:APEXresult}
\longtab[1]{
\begin{longtable}{llrrrr}
\caption{APEX results}\\ \hline\hline
Source&Molecule	&Line peak ($T_{\rm A,peak}$)\tablefootmark{b}	&$I$ &$\Delta v$&rms \\
   &	&(mK)	&(K kms$^{-1})$	&(kms$^{-1})$)	&(mK)\\ \hline
\endfirsthead
\caption{continued.}\\
\hline\hline
Source		&Molecule	&Line peak ($T_{\rm A,peak}$)\tablefootmark{b}	&$I$ &$\Delta v$ &rms\\
	&	&(mK)	&(K kms$^{-1})$	&(kms$^{-1})$)	&(mK)\\  \hline
\endhead
\hline
\endfoot
NGC~3256	&CH$_3$OH	&$>$2.1$\pm$0.2								&$>$0.06$\pm$0.03&238$\pm$43 &0.20\\
&H$_{2}$S		&1.2$\pm$0.7								&0.28$\pm$0.03	&216$\pm$61		&0.20\\
&HCN		&16.9$\pm$0.4								&2.26$\pm$0.06	&182$\pm$7		&0.38\\
&HCO$^{+}$	&21.6$\pm$0.4								&4.17$\pm$0.06	&186$\pm$4		&0.38\\
&SO			&1.2$\pm$0.4								&0.13$\pm$0.03	&85$\pm$4		&0.38\\
&HNC		&4.4$\pm$2.0								&$>$1.02$\pm$0.04&284$\pm$145	&0.38\\
\hline
NGC~1068	&CH$_3$OH	&$>$0.7$\pm$0.2								&$>$0.18$\pm$0.04	&303$\pm$185	&0.35\\
&H$_{2}$S		&1.4$\pm$0.3								&0.40$\pm$0.06		&267$\pm$60		&0.35\\
&HCN		&28.4$\pm$0.6								&7.41$\pm$0.10		&253$\pm$6		&0.60\\
&HCO$^{+}$	&20.7$\pm$0.6								&5.32$\pm$0.11		&250$\pm$8		&0.60\\
&SO			&1.0$\pm$0.5								&0.14$\pm$0.04		&133$\pm$73		&0.60\\
&HNC		&7.6$\pm$0.9								&$>$0.73$\pm$0.06	&104$\pm$18		&0.60\\
\hline
IC~860		&CH$_3$CN	&1.4$\pm$0.5								&0.39$\pm$0.09		&192$\pm$87		&0.52\\
&HCN		&2.2$\pm$0.3								&0.82$\pm$0.15		&357$\pm$63		&0.80\\
\hline
NGC~4418	&CH$_3$CN	&1.3$\pm$0.3								&0.27$\pm$0.11		&216$\pm$56		&0.41\\
&H$_{2}$S		&1.6$\pm$0.4								&0.25$\pm$0.11		&126$\pm$39		&0.41\\
&H$_2$CS	&1.7$\pm$0.5								&$>$0.11$\pm$0.03	&---			&0.41\\
&HCN		&6.4$\pm$0.4								&1.45$\pm$0.10		&200$\pm$16		&0.67\\
&HCO$^{+}$	&5.3$\pm$0.5								&0.84$\pm$0.08		&151$\pm$15		&0.67\\
&SO			&0.9$\pm$0.5								&0.18$\pm$0.07		&166$\pm$15		&0.67\\
&HNC		&3.3$\pm$0.5								&$>$0.70$\pm$0.07	&251$\pm$103	&0.67\\
\hline
Circinus	&CH$_3$OH	&$>$2.0$\pm$1.0								&$>$0.16$\pm$0.04	&63$\pm$37		&0.37\\
&H$_{2}$S		&3.0$\pm$0.4								&0.78$\pm$0.06		&241$\pm$33		&0.37\\
&HCN		&24.0$\pm$1.0								&6.0$\pm$0.1		&233$\pm$11		&0.81\\
&HCO$^{+}$	&29.5$\pm$1.0								&7.0$\pm$0.1		&238$\pm$9		&0.81\\
&SO			&4.2$\pm$1.0								&0.16$\pm$0.08		&38$\pm$9		&0.81\\
&HNC		&---										&>0.93$\pm$0.08		&---			&0.81\\
\hline
NGC~1097	&CH$_3$OH	&$>$0.8$\pm$0.2								&$>$0.04$\pm$0.04	&18$\pm$273		&0.30\\
&H$_{2}$S		&0.5$\pm$0.3								&0.11$\pm$0.04		&157$\pm$89		&0.30\\
&HCN		&5.9$\pm$0.3								&2.38$\pm$0.10		&409$\pm$26		&0.51\\
&HCO$^{+}$	&4.9$\pm$0.3								&1.97$\pm$0.09		&405$\pm$29		&0.51\\
&HNC		&2.4$\pm$1.0								&$>$0.33$\pm$0.05	&126$\pm$63		&0.51\\	
\hline
NGC~4945	&CH$_3$OH	&7.2$\pm$1.1								&$>$1.92$\pm$0.08	&238$\pm$43		&0.46\\
&CH$_3$CN	&2.8$\pm$0.3								&0.96$\pm$0.08		&345$\pm$39		&0.46\\
&H$_{2}$S		&4.8$\pm$0.5								&1.49$\pm$0.07		&295$\pm$35		&0.46\\
&H$_2$CS	&1.4$\pm$0.5								&0.46$\pm$0.07		&275$\pm$38		&0.46\\
&HCN		&95.5$\pm$7.1								&31.3$\pm$0.2		&325$\pm$28		&1.0\\
&HCO$^{+}$	&85.6$\pm$5.4								&30.1$\pm$0.2		&337$\pm$25		&1.0\\
&SO			&---										&0.24$\pm$0.16		&---			&1.0\\
&HOC$^{+}$	&5.5$\pm$1.0								&1.75$\pm$0.16		&322$\pm$69		&1.0\\	
&HNC		&38.3$\pm$2.8								&13.3$\pm$0.2		&339$\pm$28		&1.0\\					
\hline
NGC~253		&CH$_3$OH	&9.5$\pm$1.4								&$>$1.93$\pm$0.07	&183$\pm$45		&0.48\\
&CH$_3$CN	&4.7$\pm$1.0								&1.05$\pm$0.08		&208$\pm$35		&0.48\\
&H$_{2}$S		&10.4$\pm$1.1								&3.09$\pm$0.06		&187$\pm$24		&0.48\\
&H$_2$CS	&0.7$\pm$1.1								&0.12$\pm$0.05		&126$\pm$23		&0.48\\
&HCN		&161.4$\pm$3.7								&33.5$\pm$0.14		&203$\pm$5		&0.87\\
&HCO$^{+}$	&165.3$\pm$2.8								&33.7$\pm$0.14		&197$\pm$4		&0.87\\
&SO			&6.2$\pm$0.6								&1.00$\pm$0.10		&162$\pm$20		&0.87\\
&HOC$^{+}$	&6.1$\pm$0.9								&1.03$\pm$0.11		&157$\pm$27		&0.87\\	
&HNC		&$>$42.9$\pm$0.8							&$>$3.91$\pm$0.10	&135$\pm$5		&0.87\\					
\hline
NGC~1377	&HCN		&1.3$\pm$0.4								&0.76$\pm$0.1		&135$\pm$45		&0.60\\
\hline
M 83		&CH$_3$OH	&0.9$\pm$0.3								&$>$0.29$\pm$0.03	&377$\pm$242	&0.51\\
&H$_{2}$S		&2.6$\pm$0.4								&0.31$\pm$0.05		&107$\pm$19		&0.51\\
&HCN		&34.0$\pm$0.9								&3.95$\pm$0.13		&106$\pm$3		&0.96\\
&HCO$^{+}$	&36.0$\pm$0.9								&4.21$\pm$0.13		&108$\pm$3		&0.96\\
&HOC$^{+}$	&1.1$\pm$0.9								&0.39$\pm$0.14		&273$\pm$3		&0.96\\					
\hline
NGC~4826	&H$_{2}$S		&0.9$\pm$0.2								&0.27$\pm$0.09		&288$\pm$58		&0.27\\
&HCN		&5.5$\pm$0.6								&1.54$\pm$0.11		&273$\pm$31		&0.66\\
&HCO$^{+}$	&4.9$\pm$0.4								&1.38$\pm$0.13		&270$\pm$22		&0.66\\
NGC~5128 &-&-&-&-&-\\
\hline									
\end{longtable}
\tablefoot{Integrated line intensities are calculated using the formula: $I_{\rm{line}}=\sum_{\rm i} {I_{\rm i} dv}$, where $I$ is the intensity at each velocity and $dv$ is the velocity resolution.The errors are calculated using,$\Delta I_{\rm line}= \sqrt{(\Delta I_{\rm L})^2 + (\Delta I_{\rm B})^2}= \sqrt{(\sigma v_{\rm res}\sqrt{N_{\rm L}})^2+(\sigma v_{\rm res}N_{\rm L}/\sqrt{N_{\rm B}})^2}$, where $\sigma$ is the rms, $v_{\rm res}$ is the velocity resolution, $N_{\rm L}$ is the number of channels that contribute to the line and $N_{\rm B}$ is the number of channels which is outside the line. The RMS noise levels are given in Table \ref{tab:Basic}. NGC~5128 has strong continuum emission and its noise levels could not be measured.}
}


\section{Notes on Critical Densities}\label{append2}

In this paper, the critical densities of H$_{2}$S, HCN and HCO$^+$ were calculated assuming the molecules are two-level systems,

\begin{equation}
    n_{crit} = A_{ul} /q_{ul}
\end{equation}

where $A_{ul}$ is the spontaneous transition probability of the transition u to l and $q_{ul}$ is the corresponding collisional rate coefficient. This simple approach is sufficient for the comparison of critical densities in Sec.\ref{s:column}.

A more realistic estimate of the critical density is,

\begin{equation}
    n_{crit} = A_{ul} / (\sum_{l<u} q_{ul} + \sum_{u'>u} q_{uu'})
\end{equation}

where one allows for transitions between more levels. In general it gives lower values. Furthermore, opacity will also impact the critical density. For further discussion, see for example \citet{shirley2015a}.

\end{appendix}

\end{document}